\documentclass[aps,pra,twocolumn,superscriptaddress,citeautoscript,longbibliography]{revtex4-2}
\usepackage[dvipsnames]{xcolor}
\usepackage{amsmath,amssymb}
\usepackage{amsthm}
\usepackage{physics}
\usepackage[colorlinks = true,
            linkcolor = blue,
            urlcolor  = blue,
            citecolor = blue,
            anchorcolor = blue]{hyperref}
\usepackage{cleveref}
\usepackage{graphicx}
\usepackage{dsfont}
\usepackage{epstopdf}
\usepackage{bm}
\usepackage{bbm}
\usepackage{todonotes}
\usepackage{placeins}
\usepackage[normalem]{ulem}
\usepackage{CJKutf8,orcidlink}
\usepackage{algorithm}
\usepackage{algorithmic}
\usepackage{changes}

\usepackage{longtable}

\begin{document}
\begin{CJK*}{UTF8}{gbsn} 
\crefname{equation}{Eq.}{Eqs.}
\crefname{figure}{Fig.}{Fig.}
\crefname{appendix}{Appendix}{Appendix}

\title{Optimizing and Certifying Multipartite Permutationally Invariant Bell Inequalities}

\author{Jin-Fu Chen (陈劲夫)\orcidlink{0000-0002-7207-969X}}
\email{jinfuchen@lorentz.leidenuniv.nl}
\affiliation{Instituut-Lorentz, Universiteit Leiden, P.O. Box 9506, 2300 RA Leiden, The Netherlands}
\affiliation{$\langle aQa^L\rangle$ Applied Quantum Algorithms, Universiteit Leiden, The Netherlands}
\author{Mengyao Hu (胡梦瑶)
\orcidlink{0000-0003-2621-3365}}
\email{mengyao@lorentz.leidenuniv.nl}
\affiliation{Instituut-Lorentz, Universiteit Leiden, P.O. Box 9506, 2300 RA Leiden, The Netherlands}
\affiliation{$\langle aQa^L\rangle$ Applied Quantum Algorithms, Universiteit Leiden, The Netherlands}
\affiliation{Centre for Quantum Technologies, National University of Singapore, 3 Science Drive 2, Singapore 117543, Singapore}
\author{Jordi Tura
\orcidlink{0000-0002-6123-1422}}
\email{tura@lorentz.leidenuniv.nl}
\affiliation{Instituut-Lorentz, Universiteit Leiden, P.O. Box 9506, 2300 RA Leiden, The Netherlands}
\affiliation{$\langle aQa^L\rangle$ Applied Quantum Algorithms, Universiteit Leiden, The Netherlands}
\date{\today}
\begin{abstract}
Multipartite Bell nonlocality provides a device-independent probe of many-body
quantum correlations, but its characterization is limited by the rapid growth
of the underlying classical and quantum optimization problems.  We develop a scalable method
for constructing and certifying permutationally invariant Bell inequalities
using only one- and two-body correlators. The construction gives families of inequalities with robust quantum
violations for general $m$ measurements as the number of parties
$N$ becomes large. To improve robustness against noise, we optimize the ratio of the quantum value to the classical bound for these families in the large-$N$ limit. We then certify the resulting quantum violation using semidefinite programming. For the broad class of Bell inequalities studied here, the infinite-$N$ ratios take simple rational values for finite $m$ and converge to $\coth(1)$ as $m\to\infty$. The optimized inequalities efficiently detect many-body
Bell nonlocality with collective measurements, with more measurement settings leading to stronger violations.
\end{abstract}
\maketitle
\end{CJK*}

\section{Introduction}
Quantum theory predicts correlations that challenge any classical explanation
based on locality and pre-existing elements of reality. The
Einstein--Podolsky--Rosen paradox asks whether quantum mechanics could be
completed by a local realistic description \cite{einsteinCanQuantumMechanicalDescription1935,bohrCanQuantumMechanicalDescription1935}.
John Bell showed that this debate can be formulated as experimentally testable
inequalities, whose violation demonstrates that no local hidden-variable model
can reproduce all quantum predictions
\cite{bellEinsteinPodolskyRosen1964,clauserProposedExperimentTest1969,aspectExperimentalRealizationEinsteinPodolskyRosenBohm1982,brunnerBellNonlocality2014}.
In the bipartite setting, both the structure of Bell correlations and their
experimental realization are now relatively well developed
\cite{fineHiddenVariablesJoint1982,pitowskyQuantumProbabilityQuantum1989,brunnerBellNonlocality2014}. 
In particular,
loophole-free Bell tests have established Bell nonlocality as an experimentally
robust phenomenon
\cite{hensenLoopholefreeBellInequality2015,giustinaSignificantLoopholeFreeTestBells2015,shalmStrongLoopholeFreeTest2015,rosenfeldEventReadyBellTest2017},
and Bell nonlocality has become a key resource for device-independent quantum
information processing, including quantum key distribution~\cite{acinDeviceIndependentSecurityQuantum2007, pironioDeviceindependentQuantumKey2009, vaziraniFullyDeviceIndependentQuantum2014, xuSecureQuantumKey2020}, certified randomness generation~\cite{pironioRandomNumbersCertified2010,fyrillasCertifiedRandomnessTight2024,zhangRandomnessNonlocalityMultipleInput2025}, and quantum system verification through self-testing~\cite{supicSelftestingQuantumSystems2020b,liNecessarySufficientCondition2025,baccariScalableBellInequalities2020}.
These developments motivate the search for analogous certification tools in systems involving many parties.

Multipartite nonlocality is especially relevant for quantum many-body systems,
where correlations are distributed among many parties and cannot be reduced to
pairwise effects
\cite{merminExtremeQuantumEntanglement1990,svetlichnyDistinguishingThreebodyTwobody1987,bancalDeviceIndependentWitnessesGenuine2011}.
Multipartite Bell inequalities can therefore serve as witnesses of intrinsically
nonclassical properties of many-body states.
However, the multipartite setting is
substantially more challenging than the bipartite one: the number of
deterministic strategies and the complexity of the local polytope grow exponentially
with the number of parties
\cite{pitowskyQuantumProbabilityQuantum1989,pitowskyCorrelationPolytopesTheir1991,zukowskiBellsTheoremGeneral2002,bancalDetectingGenuineMultipartite2011}.
Consequently, constructing useful multipartite Bell inequalities and certifying
their classical and quantum values quickly becomes difficult
\cite{wernerAllmultipartiteBellcorrelationInequalities2001,collinsBellInequalitiesArbitrarily2002a}.

In multipartite scenarios, symmetry provides a natural way to reduce the complexity. A representative example is permutationally invariant (PI) Bell inequalities~\cite{turaDetectingNonlocalityManybody2014c,turaNonlocalityManybodyQuantum2015,wagnerBellCorrelationsManyBody2017,muller-rigatInferringNonlinearManyBody2021,guoDetectingBellCorrelations2023a}. By enforcing invariance under party relabeling and restricting to few-body correlators, the description of
correlations is projected onto a much lower-dimensional space. 
These PI Bell inequalities provide a way to certify the presence of Bell correlations in quantum many-body states directly from
experimentally accessible measurements~\cite{schmiedBellCorrelationsBoseEinstein2016,engelsenBellCorrelationsSpinSqueezed2017},
without requiring full state tomography.
However,
a systematic framework for finding PI Bell inequalities with large quantum
violations at large party number has not yet been established.

In this work, we develop a framework for both optimizing and certifying
$N$-partite  PI Bell inequalities with one- and two-body correlators, which can be extended naturally to higher-body correlators. On the
classical side, permutational invariance reduces the dimension of the
local polytope. We show the classical bound can be efficiently found by
dynamic programming~\cite{huTropicalContractionTensor2026a,huCharacterizingTranslationinvariantBell2026a}. On the quantum side, the PI Bell operators are decomposed into total-spin sectors, and we focus on the quantum violation in the symmetric subspace~\cite{turaNonlocalityManybodyQuantum2015}. The spin-squeezed state then enables quantum violation at finite $N$. At large $N$, the Holstein--Primakoff approximation \cite{fadelBellCorrelationsFinite2018,holsteinFieldDependenceIntrinsic1940} maps the large spin to a single bosonic mode. We further optimize the coefficients of PI Bell inequalities by maximizing the ratio between the quantum value and the classical bound \cite{chenOptimizingQuantumViolation2026}.

To turn these variationally found violations into rigorous statements, we
construct certificates of the quantum violation. Semidefinite relaxations based on the
Navascu\'es--Pironio--Ac\'in hierarchy provide general tools for bounding
quantum correlations and quantifying the maximal violation of a given Bell inequality
\cite{navascuesBoundingSetQuantum2007,navascuesConvergentHierarchySemidefinite2008,tavakoliSemidefiniteProgrammingRelaxations2024}. Previously, an SDP hierarchy was proposed to characterize classical correlations in PI Bell inequalities \cite{fadelBoundingSetClassical2017a}.
Here, we instead use moment relaxations and sum-of-squares decompositions to certify quantum bounds for PI Bell inequalities. For the inequalities constructed in this work, these
certificates agree with the variational values, making the quantum-to-classical ratio exact to the leading order in $N$.

The rest of the manuscript is organized as follows.
Section~\ref{sec:Nm2} introduces PI Bell inequalities in the
$(N,m,2)$ scenario with up to two-body correlators.
Section~\ref{sec:Variational approach} presents the large-$N$ variational
approach and the Holstein--Primakoff approximation.
Section~\ref{sec:finite_N_results} studies the finite-$N$ behavior and
compares exact symmetric-sector calculations with the asymptotic predictions. We conclude in
Section~\ref{sec:conclusion}.

\section{Preliminaries}
\label{sec:Nm2}

We consider the $(N,m,2)$ Bell scenario, where $N$ parties each choose one of
$m$ dichotomic measurements. The observable corresponding to setting
$k\in\{0,\ldots,m-1\}$ at site $i\in\{1,\ldots,N\}$ is denoted by
$A_k^{(i)}$, with outcomes $A_k^{(i)}=\pm1$.

A general linear functional written in the correlator form is
\begin{align}
I
=
\sum_{\ell=1}^{N}
\sum_{\substack{
1\le i_1<\cdots<i_\ell\le N\\
k_1,\ldots,k_\ell=0,\ldots,m-1
}}
\alpha^{\,i_1,\ldots,i_\ell}_{k_1,\ldots,k_\ell}
\left\langle
A_{k_1}^{(i_1)}\cdots A_{k_\ell}^{(i_\ell)}
\right\rangle .
\end{align}
In this work, we focus on PI Bell inequalities, for which the coefficients are independent of the choice of sites
\begin{align}
\alpha^{\,i_1,\ldots,i_\ell}_{k_1,\ldots,k_\ell}
=
\alpha_{k_1,\ldots,k_\ell}.
\end{align}
It is then useful to introduce the collective $\ell$-body correlators
\begin{align}
\mathcal{S}_{k_{1},\ldots,k_{\ell}}=\sum_{(i_{1},\ldots,i_{\ell})}A_{k_{1}}^{(i_{1})}\cdots A_{k_{\ell}}^{(i_{\ell})},
\end{align}
where the sum is over ordered $\ell$-tuples of pairwise distinct sites, $i_{a}\neq i_{b}$ for $a\neq b$. With this convention, a PI linear functional can be written as
\begin{align}
I
=
\sum_{\ell=1}^{N}
\sum_{0\le k_1\le\cdots\le k_\ell\le m-1}
\alpha_{k_1,\ldots,k_\ell}
\frac{
\left\langle
\mathcal S_{k_1,\ldots,k_\ell}
\right\rangle
}{
\prod_{k=0}^{m-1}n_k!
},
\label{eq:PIBell}
\end{align}
where $n_k$ is the multiplicity of the setting $k$ in the multiset
$\{k_1,\ldots,k_\ell\}$. The factorial factor removes the overcounting due to repeated settings.
For example, in the $(N,2,2)$ scenario, a PI linear functional involving correlators up to three-body order reads
$I=
\alpha_0\langle\mathcal S_0\rangle
+\alpha_1\langle\mathcal S_1\rangle
+\alpha_{0,0}\langle\mathcal S_{0,0}\rangle/2!
+\alpha_{0,1}\langle\mathcal S_{0,1}\rangle
+\alpha_{1,1}\langle\mathcal S_{1,1}\rangle/2!
+\alpha_{0,0,0}\langle\mathcal S_{0,0,0}\rangle/3!
+\alpha_{0,0,1}\langle\mathcal S_{0,0,1}\rangle/2!
+\alpha_{0,1,1}\langle\mathcal S_{0,1,1}\rangle/2!
+\alpha_{1,1,1}\langle\mathcal S_{1,1,1}\rangle/3!$.

We restrict to linear functionals containing at most two-body correlators
\begin{align}
I
=
\sum_{k=0}^{m-1}\alpha_k
\left\langle\mathcal S_k\right\rangle
+
\frac12
\sum_{k,l=0}^{m-1}
\alpha_{k,l}
\left\langle\mathcal S_{k,l}\right\rangle ,
\label{eq:Iwithuptotwobody}
\end{align}
where $\alpha_{k,l}=\alpha_{l,k}$. 
To obtain robust PI Bell inequalities in the large-$N$ regime, we focus on a useful subclass in which the two-body coefficient matrix has rank one,
\begin{align}
\alpha_{k,l}=\gamma_k\gamma_l,\label{eq:rankone_alpha}
\end{align} 
which simplifies both the classical optimization and the construction of the corresponding Bell operator. Its motivation and large-$N$ relevance are discussed in Appendix~\ref{appendix:Discussionrank1}.

Table~\ref{tab:DP_comparison} summarizes the methods used to evaluate the
classical bounds and quantum values. 
Permutational invariance considerably simplifies the computation of the classical bound, since deterministic strategies can be enumerated by occupation numbers.
This structure can be further exploited using dynamic programming to evaluate the classical bound, similarly to the tensor network contraction based on tropical algebra developed for translationally invariant Bell inequalities~\cite{huTropicalContractionTensor2026a,huCharacterizingTranslationinvariantBell2026a}.

On the quantum side, permutational invariance enables symmetric measurement settings, and the Bell operator is written in terms of collective spin operators. For the rank-one parametrization, this operator takes a form similar to the Hamiltonian of the Lipkin--Meshkov--Glick (LMG) model~\cite{lipkinValidityManybodyApproximation1965a,turaDetectingNonlocalityManybody2014c,turaNonlocalityManybodyQuantum2015}, but with a non-extensive scaling in $N$. This non-extensive structure leads to Bell violations at large $N$. The asymptotic quantum violation can then be variationally determined by the spin-squeezed states.

\begin{table*}
\centering
\begin{tabular}{c|c}
\hline
Approach & Scaling measure \\
\hline
\multicolumn{2}{c}{\textbf{Classical bound $\beta_C$}} \\
\hline
Direct deterministic enumeration
&
Computational complexity: $O(2^{mN})$
\\
PI occupation enumeration
&
Computational complexity: $O(N^{2^m-1})$
\\
Rank-one dynamic programming
&
Computational complexity: $O(2^m m N^2)$
\\
\hline
\multicolumn{2}{c}{\textbf{Quantum value $\beta_Q$}} \\
\hline
Full Hilbert-space diagonalization
&
Hilbert-space dimension: $2^N$
\\
Symmetric-sector diagonalization
&
Hilbert-space dimension: $N+1$
\\
Holstein--Primakoff approximation
&
Effective single-mode bosonic problem
\\
\hline
\end{tabular}
\caption{
Computational costs for evaluating both the classical bound and the quantum value of PI Bell inequalities with $m$ measurement settings and two outcomes. For the classical bound, direct optimization over local deterministic strategies scales as $O(2^{mN})$, while permutational invariance reduces the problem to a polynomial-size occupation enumeration for fixed $m$. In the rank-one case, the dynamic-programming approach computes the exact finite-$N$ classical bound with complexity $O(MGN^2)= O(2^m m N^2)$, where $M=|\mathcal U|\le 2^m$ is the number of distinct values $u=\boldsymbol{\gamma}\cdot\boldsymbol{A}$, and $G=\sum_{k=0}^{m-1}\gamma_k$ for positive integer $\gamma_k$. For the quantum value, permutational invariance reduces the Hilbert-space dimension from $2^N$ to the symmetric spin sector of dimension $N+1$, while the Holstein--Primakoff approximation yields an effective single-mode description in the large-$N$ limit.
}
\label{tab:DP_comparison}
\end{table*}
\subsection{Local deterministic strategies and the classical bounds}
\label{sec:local deterministic strategy and classical bound}

A local deterministic strategy for a single party is specified by assigning an outcome to each of the $m$ measurements. We denote such a strategy for each party by
\begin{align}
\boldsymbol{A}=(A_0,\ldots,A_{m-1})\in\{\pm1\}^m,\label{eq:Astrategy}
\end{align}
where $A_k=\pm1$ is the assigned outcome of measurement $k$.
By permutational invariance, the superscript $(i)$ used to label the measurements can be omitted.

In a PI setting, an $N$-party deterministic configuration is
fully characterized by the occupation numbers
\begin{align}
\sum_{\boldsymbol{A}\in\{\pm1\}^{m}}N_{\boldsymbol{A}}=N,
\end{align}
where each $N_{\boldsymbol{A}}$ is a nonnegative integer counting how many parties
use the deterministic strategy $\boldsymbol{A}$. Thus, for fixed $m$ and $N$, the local polytope is generated by
all integer occupation vectors
$\{N_{\boldsymbol{A}}\}_{\boldsymbol{A}\in\{\pm1\}^m}$
satisfying the above constraint.
Using
\begin{align}
\mathcal S_{k,l}
=
\mathcal S_k\mathcal S_l-\mathcal Z_{k,l},
\qquad
\mathcal Z_{k,l}:=\sum_{i=1}^{N}A_k^{(i)}A_l^{(i)},
\end{align}
we rewrite the linear functional~\eqref{eq:Iwithuptotwobody} as 
\begin{align}
I=\sum_{k,l=0}^{m-1}\frac{\gamma_{k}\gamma_{l}}{2}\left\langle \mathcal{S}_{k}\mathcal{S}_{l}\right\rangle +\sum_{k=0}^{m-1}\alpha_{k}\left\langle \mathcal{S}_{k}\right\rangle -\sum_{k,l=0}^{m-1}\frac{\gamma_{k}\gamma_{l}}{2}\left\langle \mathcal{Z}_{k,l}\right\rangle .
\label{eq:I_rankone_collective}
\end{align}
The collective correlators under the local deterministic strategy are
\begin{align}
\begin{aligned}
\left\langle \mathcal S_k\right\rangle
=
\sum_{\boldsymbol A}
N_{\boldsymbol A}A_k,
\qquad
\left\langle \mathcal Z_{k,l}\right\rangle
=
\sum_{\boldsymbol A}
N_{\boldsymbol A}A_k A_l.
\end{aligned}
\label{eq:SkandZklwithNa}
\end{align}
Since these quantities are fixed numbers, we have $\left\langle \mathcal{S}_{k,l}\right\rangle =\left\langle \mathcal{S}_{k}\mathcal{S}_{l}\right\rangle -\left\langle \mathcal{Z}_{k,l}\right\rangle =\left\langle \mathcal{S}_{k}\right\rangle \left\langle \mathcal{S}_{l}\right\rangle -\left\langle \mathcal{Z}_{k,l}\right\rangle  $,
and Eq.~\eqref{eq:I_rankone_collective} simplifies to
\begin{align}
I=\frac{1}{2}\left(\sum_{\boldsymbol{A}}N_{\boldsymbol{A}}\,\boldsymbol{\gamma}\cdot\boldsymbol{A}\right)^{2}+\sum_{\boldsymbol{A}}N_{\boldsymbol{A}}\left[\boldsymbol{\alpha}\cdot\boldsymbol{A}-\frac{1}{2}\left(\boldsymbol{\gamma}\cdot\boldsymbol{A}\right)^{2}\right],
\label{eq:I_Na_formgamma}
\end{align}
where $\boldsymbol\alpha=(\alpha_0,\ldots,\alpha_{m-1})$ and $\boldsymbol\gamma=(\gamma_0,\ldots,\gamma_{m-1})$.
In the following, $\boldsymbol\alpha$ denotes the one-body coefficients, while
the two-body coefficients are fixed by $\alpha_{k,l}=\gamma_k\gamma_l$.
Equation~\eqref{eq:I_Na_formgamma} is well suited to dynamic programming. The details of the procedure are given in Appendix~\ref{appendix:dynamicprogramming}.

The classical bound $\beta_C$
is obtained by minimizing $I$ over all admissible occupation vectors $N_{\boldsymbol{A}}$, i.e.,
\begin{align}
    \beta_{C}=\min_{\{N_{\boldsymbol{A}}\}}I.
\end{align}
We show the calculation of the classical bounds in the thermodynamic limit $N\to\infty$ in Appendix~\ref{appendix:classicalbound}.

\subsection{Bell operator and quantum values}
\label{sec:Bell operator and quantum value}

We obtain the Bell operator by substituting a set of local measurements
into the linear functional in Eq.~\eqref{eq:Iwithuptotwobody}. We assume that all parties share the same measurements and restrict the state to the symmetric subspace. The
local observables are chosen in the $x$-$z$ plane, i.e.,
\begin{align}
\hat A_k^{(i)}
=
\cos\theta_k\,\hat\sigma_z^{(i)}
+
\sin\theta_k\,\hat\sigma_x^{(i)},
\label{eq:local_measurements}
\end{align}
where $k=0,\ldots,m-1$ labels the measurement choice and
$\hat{\sigma}_\mu ^{(i)}, \mu \in \{x,z\}$ denotes the Pauli matrix acting on site $i$. The
quantum value $\beta_Q$ is then obtained by optimizing over the angles
$\boldsymbol\theta=(\theta_0,\ldots,\theta_{m-1})$ and symmetric
many-body states.

In Eq.~\eqref{eq:Iwithuptotwobody}, the one-body collective observables are
\begin{align}
\hat{\mathcal S}_k
=
\sum_{i=1}^{N}\hat A_k^{(i)}
=
2\cos\theta_k\,\hat S_z
+
2\sin\theta_k\,\hat S_x,
\label{eq:Shatk}
\end{align}
where
\begin{align}
\hat S_\mu
=
\frac12\sum_{i=1}^{N}\hat\sigma_\mu^{(i)},
\qquad
\mu=x,y,z.
\end{align}
Here we restrict the variational optimization to the symmetric subspace of $N$
qubits, corresponding to the spin-$N/2$ irreducible representation of
$\mathrm{SU}(2)$. 
Note that this is a restriction of the full Hilbert space, the
minimum obtained in this subspace gives a variational upper bound on the quantum value.
We use a sum-of-squares certificate to obtain a certified lower bound on the quantum value (see Appendix~\ref{appendix:certification_approach}).
In the $\hat S_z$ eigenbasis $\{|M\rangle\}$, with
$M=N/2,N/2-1,\ldots,-N/2$, the relevant collective spin matrices are
\begin{align}
\begin{aligned}
(\hat{S}_{x})_{M,M'}&=\frac{1}{2}\sqrt{(\frac{N}{2}-M')(\frac{N}{2}+M'+1)}\delta_{M,M'+1}\\&+\frac{1}{2}\sqrt{(\frac{N}{2}+M')(\frac{N}{2}-M'+1)}\delta_{M,M'-1},\\(\hat{S}_{y})_{M,M'}&=-\frac{i}{2}\sqrt{\left(\frac{N}{2}-M'\right)\left(\frac{N}{2}+M'+1\right)}\,\delta_{M,M'+1}\\&+\frac{i}{2}\sqrt{\left(\frac{N}{2}+M'\right)\left(\frac{N}{2}-M'+1\right)}\,\delta_{M,M'-1},\\(\hat{S}_{z})_{M,M'}&=M\delta_{M,M'}.
\end{aligned}
\end{align}

The two-body collective correlator is defined by excluding same-site terms. 
For quantum observables, the operators acting on the same site generally do not commute. 
We use the operator
\begin{align}
\hat{\mathcal Z}_{k,l}
:=
\frac12
\sum_{i=1}^{N}
\left\{
\hat A_k^{(i)},\hat A_l^{(i)}
\right\}.
\label{eq:Zkl_quantum_def}
\end{align}
Using Eq.~\eqref{eq:local_measurements}, one obtains
\begin{align}
\hat{\mathcal Z}_{k,l}
=
N\cos(\theta_k-\theta_l),
\label{eq:hatZkl}
\end{align}
and the two-body correlator becomes
\begin{align}
\hat{\mathcal{S}}_{k,l}=\frac{1}{2}\{\hat{\mathcal{S}}_{k},\hat{\mathcal{S}}_{l}\}-\hat{\mathcal{Z}}_{k,l}.
\end{align}
Thus, the Hermitian Bell operator associated with Eq.~\eqref{eq:Iwithuptotwobody} can be written as
\begin{align}
\hat{I}(\boldsymbol{\theta})=\sum_{k=0}^{m-1}\alpha_{k}\hat{\mathcal{S}}_{k}+\frac{1}{2}(\sum_{k=0}^{m-1}\gamma_{k}\hat{\mathcal{S}}_{k})^{2}-\sum_{k,l=0}^{m-1}\frac{\gamma_{k}\gamma_{l}}{2}\hat{\mathcal{Z}}_{k,l}.
\label{eq:Bell_operator_I_theta}
\end{align}
For fixed coefficients $\boldsymbol{\alpha}$ and $\boldsymbol{\gamma}$, the variational quantum value in the symmetric subspace is
\begin{align}
\beta_Q
=
\min_{\boldsymbol\theta,\ket\psi}
\bra\psi
\hat I(\boldsymbol\theta)
\ket\psi,\label{eq:betaqexact}
\end{align}
where $\ket\psi$ is restricted to the symmetric subspace.

Since all local measurements lie in the $x$-$z$ plane, the Bell operator has the general quadratic collective-spin form
\begin{align}
\hat I(\boldsymbol\theta)
=
J
+
J_x\hat S_x
+
J_z\hat S_z
+
\frac12
\left(
\gamma_x\hat S_x+\gamma_z\hat S_z
\right)^2,
\label{eq:bell_operator_rankone}
\end{align}
with the coefficients
$J_x=2\sum_{k=0}^{m-1}\alpha_k\sin\theta_k$,
$J_z=2\sum_{k=0}^{m-1}\alpha_k\cos\theta_k$,
$\gamma_x=2\sum_{k=0}^{m-1}\gamma_k\sin\theta_k$,
$\gamma_z=2\sum_{k=0}^{m-1}\gamma_k\cos\theta_k$, and
$J=-N(\gamma_{x}^{2}+\gamma_{z}^{2})/8$.
Therefore, in the rank-one case, the Bell operator has an LMG-like
collective-spin structure. However, the two-body coefficients $J_{x,x}$,
$J_{x,z}$, and $J_{z,z}$ have a different $N$-scaling compared to those of the LMG model~\cite{lipkinValidityManybodyApproximation1965a}. This non-extensive structure is crucial for Bell violations in the large-$N$ regime, as discussed in
Appendix~\ref{appendix:Discussionrank1}. In this regime, the asymptotic
quantum violation can be obtained variationally with spin-squeezed states.

\section{Large-$N$ Variational method}\label{sec:Variational approach}
We now describe a variational method for evaluating the quantum value using spin-squeezed states in the symmetric subspace. In the large-$N$ limit, the ground state of the Hamiltonian corresponding to the Bell operator in Eq.~\eqref{eq:bell_operator_rankone} can be well approximated by a spin-squeezed state~\cite{maQuantumSpinSqueezing2011a}. We then derive an asymptotic analytical expression for the quantum value using the Holstein--Primakoff transformation~\cite{holsteinFieldDependenceIntrinsic1940}.

\subsection{Spin-squeezed states}
\label{sec:spinsqueezedstate}

In the symmetric subspace, the expectation value of
the Bell operator in Eq.~\eqref{eq:bell_operator_rankone} is
\begin{align}
\begin{aligned}
\langle\hat I(\boldsymbol\theta)\rangle
&=
J+J_xS_x+J_zS_z
+\frac12(\gamma_xS_x+\gamma_zS_z)^2
\\&+\frac12
\bigl(
\gamma_x^2\Gamma_{x,x}
+2\gamma_x\gamma_z\Gamma_{x,z}
+\gamma_z^2\Gamma_{z,z}
\bigr),
\end{aligned}
\label{eq:I_spin_squeezed}
\end{align}
where the first moments are
\begin{align}
S_{\mu}=\langle\hat{S}_{\mu}\rangle=\bra{\psi}\hat{S}_{\mu}\ket{\psi},\qquad\mu=x,y,z,\label{eq:Smumean}
\end{align}
and the covariance terms are
\begin{align}
\Gamma_{\mu,\nu}
=
\frac12
\langle
\hat S_\mu\hat S_\nu+\hat S_\nu\hat S_\mu
\rangle
-
\langle\hat S_\mu\rangle
\langle\hat S_\nu\rangle.\label{eq:Gammamunucovariance}
\end{align}

To obtain a variational upper bound on the quantum value, we consider a
spin-squeezed state parametrized as
\begin{align}
|\psi_{\mathrm{sq}}(\phi,\chi)\rangle
=
e^{-i\phi \hat S_y}
e^{-i\frac{\chi}{2}
(\hat S_x\hat S_y+\hat S_y\hat S_x)}
\left|S,S\right\rangle,
\label{eq:spin_squeezed_ansatz}
\end{align}
with
$S=N/2$. The
state $|S,S\rangle$ is the fully polarized Dicke state satisfying
$\hat S_z|S,S\rangle=S|S,S\rangle$.
Here $\phi$ sets the mean-spin direction in the $x\text{-}z$ plane, and
$\chi$ controls the squeezing of the transverse fluctuations. Together, these parameters optimize the mean-field and fluctuation contributions in Eq.~\eqref{eq:I_spin_squeezed}. 
The expectation value over the spin-squeezed state is then
\begin{align}
E_{\mathrm{sq}}(\boldsymbol\theta,\phi,\chi)
=
\bra{\psi_{\mathrm{sq}}(\phi,\chi)}
\hat I(\boldsymbol\theta)
\ket{\psi_{\mathrm{sq}}(\phi,\chi)} .
\label{eq:spin_squeezed_variational_energy}
\end{align}
A variational upper bound is obtained by
minimizing over both the measurement angles and the state parameters:
\begin{align}
\beta_Q^{\mathrm{sq}}
=
\min_{\boldsymbol\theta,\phi,\chi}
E_{\mathrm{sq}}(\boldsymbol\theta,\phi,\chi).
\label{eq:spin_squeezed_variational_bound}
\end{align}
Since the minimization is performed over a restricted family of states, one has
\begin{align}
\beta_Q
\le
\beta_Q^{\mathrm{sq}},
\end{align}
where $\beta_Q$ is given by Eq.~\eqref{eq:betaqexact}. 

We analyze the $N$-scaling in Eq.~\eqref{eq:I_spin_squeezed}. The linear term
$J+J_xS_x+J_zS_z$ is $O(N)$. The rank-one mean-field term
$(\gamma_xS_x+\gamma_zS_z)^2$ is generically $O(N^2)$, but the optimized
mean spin suppresses it so that
$(\gamma_xS_x+\gamma_zS_z)^2=o(N)$. Spin squeezing makes the relevant variance
subextensive,
$
\gamma_x^2\Gamma_{x,x}
+2\gamma_x\gamma_z\Gamma_{x,z}
+\gamma_z^2\Gamma_{z,z}
=o(N).
$
In the thermodynamic limit $N\to\infty$, the covariance terms can therefore be neglected.

\subsection{Holstein--Primakoff transformation}

For large $N$, quantum fluctuations around a semiclassical spin direction in the symmetric subspace ($S=N/2$) can be mapped to a single bosonic mode through the Holstein--Primakoff transformation \cite{holsteinFieldDependenceIntrinsic1940}, which gives an asymptotic estimate of the quantum value.

Starting from Eq.~\eqref{eq:bell_operator_rankone}, we choose a coordinate system in the
$x\text{-}z$ plane such that one axis is aligned with the direction
$(\gamma_x,\gamma_z)$. Let
\begin{align}
    \gamma=\sqrt{\gamma_x^2+\gamma_z^2},
\qquad
\gamma_x=\gamma\sin\varphi,
\qquad
\gamma_z=\gamma\cos\varphi .\label{eq:gammaassquares}
\end{align}
We define the spin component along this direction as
\begin{align}
\hat S_\gamma
=
\sin\varphi\,\hat S_x
+
\cos\varphi\,\hat S_z .
\label{eq:S_gamma_definition}
\end{align}
We also introduce a transverse component $\hat S_\perp$, orthogonal to
$\hat S_\gamma$ in the $x$-$z$ plane. Since the transverse direction is
defined only up to a sign, we choose this sign such that the coefficient of
$\hat S_\perp$ is nonnegative. With this convention, the linear
spin term decomposes as
\begin{align}
J_x\hat S_x+J_z\hat S_z
=
J_\gamma\hat S_\gamma
+
J_\perp\hat S_\perp ,
\label{eq:linear_field_gamma_perp_decomposition}
\end{align}
where
\begin{align}
J_\gamma
=
\frac{J_x\gamma_x+J_z\gamma_z}
{\sqrt{\gamma_x^2+\gamma_z^2}},
\qquad
J_\perp
=
\frac{\left|J_x\gamma_z-J_z\gamma_x\right|}
{\sqrt{\gamma_x^2+\gamma_z^2}} .
\label{eq:J_gamma_J_perp_definitions}
\end{align}
With this convention, we assume $\gamma>0$ and $J_\perp>0$, and Eq.~\eqref{eq:bell_operator_rankone}
becomes
\begin{align}
\hat I(\boldsymbol{\theta})
=
J
+
J_\gamma \hat S_\gamma
+
J_\perp \hat S_\perp
+
\frac12\gamma^2 \hat S_\gamma^2 .
\label{eq:bell_operator_rotated_gamma_perp}
\end{align}
The quadratic term suppresses a macroscopic component along $\hat S_\gamma$.
Therefore, at leading order, the classical spin lies close to the transverse
direction. The sign convention above then selects the direction
$-\hat S_\perp$. When $J_\gamma\neq0$, the optimum is shifted slightly along
the $\gamma$ direction. In the Holstein–Primakoff description this shift is
captured by the displacement of the bosonic quadrature.

We choose the quantization axis along $-\hat S_\perp$. The spin operators are
represented as
\begin{align}
\hat S_\perp
&=
-S+\hat a^\dagger\hat a,
\label{eq:exact_HP_S_vert}\\
\hat S_\gamma
&=
\frac{1}{2}
\left[
\sqrt{2S-\hat a^\dagger\hat a}\,\hat a
+
\hat a^\dagger
\sqrt{2S-\hat a^\dagger\hat a}
\right],\label{eq:exact_HP_Sperp_Sgamma}
\end{align}
where $\hat{a}$ and $\hat{a}^\dagger$ are the bosonic annihilation and creation operators acting on the Fock space satisfying the commutation relation $[\hat{a},\hat{a}^\dagger ]=1$.
Expanding $\sqrt{2S-\hat a^\dagger\hat a}$ in powers of $\hat a^\dagger\hat a/(2S)$, one obtains
\begin{align}
\hat S_\gamma
=
\sqrt{\frac{S}{2}}
\left(
\hat a+\hat a^\dagger
\right)
+O(S^{-1/2}).
\label{eq:HP_leading_Sgamma}
\end{align}
Substituting Eqs.~\eqref{eq:exact_HP_S_vert} and~\eqref{eq:HP_leading_Sgamma} into Eq.~\eqref{eq:bell_operator_rotated_gamma_perp}, we obtain
\begin{align}
\begin{aligned}
\hat I_{\mathrm{HP}}
&\approx
J
-
J_\perp S
+
J_\perp \hat a^\dagger\hat a
\\
&\quad
+
J_\gamma
\sqrt{\frac{S}{2}}
\left(
\hat a+\hat a^\dagger
\right)
+
\frac{\gamma^2 S}{4}
\left(
\hat a+\hat a^\dagger
\right)^2 ,
\end{aligned}
\label{eq:HPtransformofbelloperator}
\end{align}
which can be diagonalized explicitly. Introducing
\begin{align}
    \hat x=\frac{\hat a+\hat a^\dagger}{\sqrt2},
\qquad
\hat p=\frac{\hat a-\hat a^\dagger}{i\sqrt2},
\qquad
[\hat x,\hat p]=i,
\end{align}
we rewrite Eq.~\eqref{eq:HPtransformofbelloperator} as
\begin{align}
\begin{aligned}
\hat{I}_{\mathrm{HP}}&=J-J_{\perp}S-\frac{J_{\perp}}{2}-\frac{J_{\gamma}^{2}S}{2\left(J_{\perp}+\gamma^{2}S\right)}\\&+\frac{J_{\perp}}{2}\hat{p}^{2}+\frac{J_{\perp}+\gamma^{2}S}{2}\left(\hat{x}+\frac{J_{\gamma}\sqrt{S}}{J_{\perp}+\gamma^{2}S}\right)^{2},
\end{aligned}
\label{eq:HP_quadrature_completed_square}
\end{align}
which is the Hamiltonian of a harmonic oscillator with a shifted equilibrium position.
The ground-state energy within the Holstein–Primakoff approximation
is therefore
\begin{align}
\begin{aligned}
    E_{\mathrm{HP}}&=J-J_{\perp}S-\frac{J_{\perp}}{2}-\frac{J_{\gamma}^{2}S}{2\left(J_{\perp}+\gamma^{2}S\right)}\\&+\frac{1}{2}\sqrt{J_{\perp}\left(J_{\perp}+\gamma^{2}S\right)},
\end{aligned}
\label{eq:HPgroundstateenergy}
\end{align}
which is the ground-state energy of the harmonic oscillator. 
\section{Optimized Bell Inequalities and Finite-\texorpdfstring{$N$}{N} Scaling}
\label{sec:finite_N_results}
In Appendix~\ref{appendix:optimizingratioatinfiniteN}, we derive robust PI Bell inequalities by optimizing their quantum-to-classical ratios. Here we study the finite-$N$ quantum violations for $m=2,\ldots,6$ measurement settings. For each Bell inequality, we compute the finite-$N$ classical bound $\beta_C$ by dynamic programming over deterministic local strategies, as described in Appendix~\ref{appendix:dynamicprogramming}, and obtain the quantum value $\beta_Q$ by optimizing the measurement angles $\boldsymbol{\theta}$ and diagonalizing the Bell operator $\hat I(\boldsymbol{\theta})$ in the symmetric spin-$N/2$ subspace.
We define
the finite-$N$ quantum-to-classical ratio~\cite{chenOptimizingQuantumViolation2026} as 
\begin{align}
    \Delta_{N,m}
    :=
    \frac{\beta_Q}{\beta_C}.
\end{align}
For fixed coefficients, the infinite-$N$ classical bound and quantum value are derived in Appendices~\ref{appendix:classicalbound} and~\ref{appendix:quantumvalue}, respectively. The optimization of the asymptotic ratio $\Delta_{\infty,m}$ over the coefficients is described in Appendix~\ref{appendix:optimizingratioatinfiniteN}. 
We show the classical bounds and asymptotic quantum values for these optimized PI Bell inequalities in the following.

\subsection{Optimized Bell inequalities with asymptotic quantum values}
\label{subsec:finite_N_classical_bounds}

\begin{figure}
    \centering
    \includegraphics[width=\linewidth]{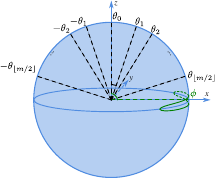}
    \caption{Diagram for the measurement angles and the spin-squeezed state. All measurement directions lie in the $x$-$z$ plane. Under the paired structure of $\boldsymbol{\alpha}$ and $\boldsymbol{\gamma}$, the angles are chosen as $\theta_{j+\lfloor m/2\rfloor}=-\theta_j$, with one unpaired central measurement $\theta_0=0$ for odd $m$. The spin-squeezed state is polarized along $x$ ($\phi=\pi/2$), with reduced $z$ fluctuations and enhanced $y$ fluctuations, as shown by the green curve.
    }
    \label{fig:diagrammeasurement}
\end{figure}
To obtain a tractable optimization problem, we impose a paired structure on the coefficients and measurement angles of the linear functionals in Eq.~\eqref{eq:I_rankone_collective}. For even $m$, we introduce the paired coefficients
\begin{align}
\begin{aligned}
\boldsymbol{\gamma}
&=
(\gamma_1,\ldots,\gamma_{m/2},\gamma_1,\ldots,\gamma_{m/2}),
\\
\boldsymbol{\alpha}
&=
(-\alpha_1,\ldots,-\alpha_{m/2},\alpha_1,\ldots,\alpha_{m/2}),
\end{aligned}
\label{eq:gammaeven}
\end{align}
All local measurements are restricted to
the $x$-$z$ plane, and
the measurement angles used for the optimized inequalities are shown in
Fig.~\ref{fig:diagrammeasurement}, and are explicitly
\begin{align}
\boldsymbol{\theta}
=
(\theta_1,\ldots,\theta_{m/2},-\theta_1,\ldots,-\theta_{m/2}).\label{eq:thetaparameterize}
\end{align}
For odd $m$, we place the unpaired component first and divide the remaining components into $(m-1)/2$ pairs; the corresponding paired coefficients are given in Appendix~\ref{appendix:classicalbound}.
Under the paired structure, the Bell operator \eqref{eq:bell_operator_rankone} is simplified to
\begin{align}
\hat I(\boldsymbol\theta)
=
J
+
J_x\hat S_x
+
\frac{1}{2}\gamma_z^2\hat S_z^2,
\label{eq:bell_operator_reversal_even}
\end{align}
where the coefficients are $\gamma_x=0$, $\gamma_z
=
4\sum_{j=1}^{m/2}\gamma_j\cos\theta_j$, $J_z=0$, and $J_x=
-4\sum_{j=1}^{m/2}\alpha_j\sin\theta_j$.
Its expectation value is $\langle\hat{I} (\boldsymbol{\theta})\rangle=J+J_x S_x +\gamma_z^2 (S_z ^2 +\Gamma _{z,z})/2$. The optimized
mean-spin direction suppresses the $S_z^2$ term, while spin squeezing reduces
the fluctuation contribution $\Gamma_{z,z}$. A similar reduction holds for odd $m$.

In the following, we list the optimized coefficients, the exact finite-$N$ classical bounds, and the corresponding asymptotic quantum values for $m=2,...,6$. For sufficiently large $N$, the classical bounds become linear in $N$, with small finite-size corrections for odd-$m$ cases. 

For $m=2$, we use
\begin{align}
\boldsymbol{\alpha}=(-1,1),
\qquad
\boldsymbol{\gamma}=(1,1).\label{eq:bellm=2}
\end{align}
The classical bound is
\begin{align}
\beta_C=-2N,
\qquad N\ge 2,
\end{align}
and the asymptotic quantum value is
\begin{align}
\beta_Q=-\frac{5}{2}N+o(N).
\end{align}

For $m=3$, we use
\begin{align}
\boldsymbol{\alpha}=(0,-45,45),
\qquad
\boldsymbol{\gamma}=(4,5,5).\label{eq:bellm=3}
\end{align}
The classical bound is
\begin{align}
\beta_C
=
\begin{cases}
-98N+8, & N=3,\\
-98N+2, & N=5,7,\\
-98N, & N=2,4,6\ \text{or}\ N\ge 8,
\end{cases}
\end{align}
and the asymptotic quantum value is
\begin{align}
\beta_Q=-126N+o(N).
\end{align}

For $m=4$, we use
\begin{align}
\boldsymbol{\alpha}=(-225,-931,225,931),
\quad
\boldsymbol{\gamma}=(15,19,15,19).\label{eq:bellm=4}
\end{align}
The classical bound is
\begin{align}
\beta_C=-2312N,
\qquad N\ge 2,
\end{align}
and the asymptotic quantum value is
\begin{align}
\beta_Q=-\frac{6001}{2}N+o(N).
\end{align}

For $m=5$, we use
\begin{align}
\begin{aligned}
\boldsymbol{\alpha}
&=
(0,-11700,-31525,11700,31525),
\\
\boldsymbol{\gamma}
&=
(72,78,97,78,97).
\end{aligned}\label{eq:bellm=5}
\end{align}
The classical bound is
\begin{align}
\beta_C
=
\begin{cases}
-89042N+578, & N=3,\\
-89042N+242, & N=5,\\
-89042N+8, & N=7,9,\\
-89042N+2, & N=11,13,\ldots,23,\\
-89042N, & N\ \mathrm{even}\ \text{or}\ N\ge 25,
\end{cases}
\end{align}
and the asymptotic quantum value is
\begin{align}
\beta_Q=-116050N+o(N).
\end{align}

For $m=6$, we use
\begin{align}
\begin{aligned}
\boldsymbol{\alpha}
&=
(-1500625,-5207475,-11362901,\\&\quad
1500625,5207475,11362901),
\\
\boldsymbol{\gamma}
&=
(1225,1365,1661,1225,1365,1661).
\end{aligned}\label{eq:bellm=6}
\end{align}
The classical bound is
\begin{align}
\beta_C=-36142002N,
\qquad N\ge 2,
\end{align}
and the asymptotic quantum value is
\begin{align}
\beta_Q=-\frac{94424629}{2}N+o(N).
\end{align}

The resulting asymptotic quantum-to-classical ratios $\Delta_{\infty,m}^{\mathrm{opt}}=\beta_Q/\beta_C$ of the above optimized Bell inequalities are summarized in
Table~\ref{tab:infiniteN-different-m}. 
The linear scaling coefficient of the quantum value agrees with the sum-of-squares certification presented in Appendix~\ref{appendix:certification_approach}.
The optimized ratios increase with $m$ and approach the continuum limit
$\coth(1)$, whose derivation is given in Appendix~\ref{sec:Nminftyresults}.

\begin{center}
\setlength{\LTcapwidth}{\linewidth}
\begin{longtable}{c|c|c}
    \caption{Optimized asymptotic quantum-to-classical ratios for the $(N,m,2)$ scenario in
the limit $N\to\infty$. The second column shows $\Delta_{\infty,m}$ with $\gamma_k=1$ and $\boldsymbol{\alpha}=(-m+1,-m+3,...,m-3,m-1)$ \cite{wagnerBellCorrelationsManyBody2017},
    and the third column shows the ratio $\Delta_{\infty,m}^{\mathrm{opt}}$ for the optimized Bell inequalities obtained in this work. }
    \label{tab:infiniteN-different-m} \\
    \hline
    \rule{0pt}{3mm}$m$ & $\Delta_{\infty,m}$ &$\Delta_{\infty,m}^{\mathrm{opt}}$  \\[0.3mm]
    \hline
    \endfirsthead

    \hline
    \multicolumn{3}{l}{\textit{(continued from previous page)}} \\
    \hline
    \rule{0pt}{3mm}$m$ & $\Delta_{\infty,m}$ &$\Delta_{\infty,m}^{\mathrm{opt}}$  \\[0.3mm]
    \hline
    \endhead

    \hline
    \multicolumn{3}{r}{\textit{(continued on next page)}} \\
    \endfoot

    \hline
    \endlastfoot

    2         & $5/4$              & $5/4$ \\
    3         & $1.28458$          & $9/7$ \\
    4         & $1.29711$          & $353/272$ \\
    5         & $1.30285$          & $275/211$ \\
    6         & $1.30597$          & $66637/51012$ \\
    $\infty$  & $\coth(1)$         & $\coth(1)$ \\

\end{longtable}
\end{center}

\subsection{Finite-\texorpdfstring{$N$}{N} convergence of quantum-to-classical ratios}

We show the quantum-to-classical ratios $\Delta_{N,m}$ for the above PI Bell inequalities at finite $N$ in the following. At the optimized measurement angles $\boldsymbol{\theta}^*$, we also evaluate the large-$N$ variational energy $E_{\mathrm{HP}}$ from the Holstein--Primakoff approximation in Eq.~\eqref{eq:HPgroundstateenergy}, together with a certified lower bound on the energy obtained from the moment relaxation in Eq.~\eqref{eq:f_SxSz}. Details of the certification are given in Appendix~\ref{appendix:certifyLMGenergy}.
Since the classical bound is negative $\beta_C<0$, the corresponding quantum-to-classical ratios obey
\begin{align}
    \Delta^{\mathrm{cert}}
    \ge
    \Delta^{\mathrm{exact}}
    \ge
    \Delta^{\mathrm{HP}} ,
    \label{eq:certexactHPrelation}
\end{align}
where the certified quantum value is given by Eq.~\eqref{eq:f_SxSz}, the exact quantum value is given by Eq.~\eqref{eq:betaqexact}, and Eq.~\eqref{eq:HPgroundstateenergy} gives the quantum value within the Holstein--Primakoff approximation.

\begin{figure*}
    \centering
    \includegraphics[width=\linewidth]{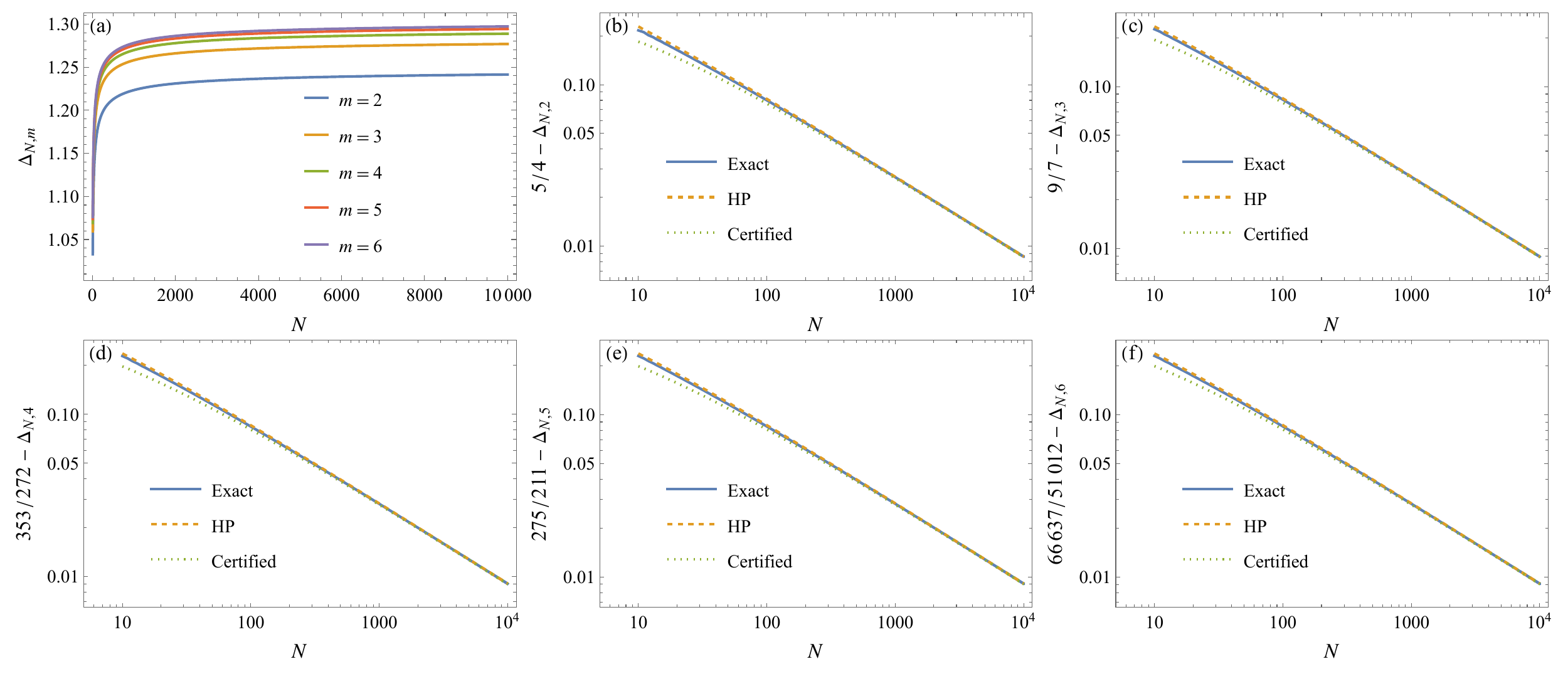}
\caption{
Finite-$N$ quantum-to-classical ratios for the optimized PI Bell inequalities.
(a) Exact ratios $\Delta_{N,m}=\beta_Q/\beta_C$ for
$m=2,\ldots,6$, obtained by optimizing the measurement angles
$\boldsymbol{\theta}$. (b)--(f) Finite-size difference
$\Delta_{\infty,m}-\Delta_{N,m}$ on a log-log scale for
$m=2,\ldots,6$, respectively. The dashed and dotted curves show the Holstein--Primakoff (HP)
approximation~\eqref{eq:HPgroundstateenergy} and the certified moment-relaxation bound~\eqref{eq:f_SxSz}, respectively. 
}
    \label{fig:ratio_all_m}
\end{figure*}

The numerical results are shown in Fig.~\ref{fig:ratio_all_m}. Figure~\ref{fig:ratio_all_m}(a) gives the exact finite-$N$ ratios $\Delta_{N,m}$ for $m=2,\ldots,6$. The ratios increase with $N$ and approach the corresponding asymptotic values $\Delta_{\infty,m}^{\mathrm{opt}}$.
Figure~\ref{fig:ratio_all_m}(b)--(f) show the finite-size difference
$\Delta_{\infty,m}^{\mathrm{opt}}-\Delta_{N,m}$ on a log-log scale for $m=2,...,6$, respectively. The difference decreases with $N$, indicating convergence of the finite-$N$ optimization to the infinite-$N$ prediction (see Table~\ref{tab:infiniteN-different-m}). We also compare the exact ratios (solid) with the Holstein–Primakoff approximation~\eqref{eq:HPgroundstateenergy} (dashed) and the certified moment-relaxation bound~\eqref{eq:f_SxSz} (dotted), both evaluated at the optimized angles $\boldsymbol{\theta}^*$. Throughout the range $10 \leq N \leq 10000$, the
ordering in Eq.~\eqref{eq:certexactHPrelation} holds.

\section{Conclusion}\label{sec:conclusion}
Multipartite Bell nonlocality can be detected by PI Bell inequalities involving only one- and two-body correlators \cite{turaDetectingNonlocalityManybody2014c,turaNonlocalityManybodyQuantum2015,wagnerBellCorrelationsManyBody2017}. In this work, we construct a family of PI Bell inequalities with dichotomic measurements and optimize their coefficients to improve the quantum-to-classical ratio. For these inequalities, spin-squeezed states give scalable pathway to probe Bell correlations in large systems, and their quantum violations remain robust to noise at large $N$. As the system size increases, the finite-$N$ ratios converge to the corresponding infinite-$N$ ratios. In the infinite-$N$ limit, the optimized ratios are rational numbers for finite $m$ and approach $\coth(1)$ as $m\to\infty$. Therefore, increasing the number of measurement settings can lead to more robust PI Bell inequalities with larger quantum-to-classical ratios.

We also certified the optimized large-$N$ violations using moment relaxations
and sum-of-squares decompositions, following semidefinite and moment-based
approaches to quantum correlations
\cite{navascuesBoundingSetQuantum2007,navascuesConvergentHierarchySemidefinite2008}.  The certified bounds
match the variational values, showing that the asymptotic scaling of the quantum value is exact to leading order in $N$.
Our results provide a practical route for detecting many-body Bell nonlocality
with collective observables. These optimized inequalities are suited
to spin-squeezed atomic ensembles and other platforms where collective
measurements are natural and scalable
\cite{schmiedBellCorrelationsBoseEinstein2016,engelsenBellCorrelationsSpinSqueezed2017}.

Several extensions can be further investigated. One direction is to go beyond one- and two-body correlators and optimize PI Bell inequalities involving higher-order correlations, as explored for detecting Bell correlations with non-Gaussian states~\cite{guoDetectingBellCorrelations2023a}. Incorporating such correlators may lead to stronger Bell violations and larger quantum-to-classical ratios. A complementary direction is to investigate PI Bell inequalities with more than two outcomes. For example, three-outcome inequalities are suited to spin-1 systems and give scalable local-dimension witnesses~\cite{muller-rigatThreeoutcomeMultipartiteBell2026}. 
More generally, both higher-order and multi-outcome scenarios pose additional challenges for numerical certification, as they lead to larger moment matrices and may require stronger SDP relaxations. Achieving scalability will therefore likely rely on symmetry-adapted bases and moment-matrix representations that explicitly exploit permutational invariance.

\section*{Acknowledgements}
We thank Pawe{\l} Cie\'sli\'nski and Jiajie Guo for the helpful discussion. 
J.F.C. and J.T. acknowledge the support received from the European Union's Horizon Europe research and innovation programme through the ERC StG FINE-TEA-SQUAD (Grant No.~101040729). 
This publication is part of the `Quantum Inspire - the Dutch Quantum Computer in the Cloud' project (with project number [NWA.1292.19.194]) of the NWA research program `Research on Routes by Consortia (ORC)', which is funded by the Netherlands Organization for Scientific Research (NWO). 
This project is supported by the National Research Foundation, Singapore through the National Quantum Office, hosted in A*STAR, under its Centre for Quantum Technologies Funding Initiative (S24Q2d0009).
Parts of this work were performed by using the compute
resources from the Academic Leiden Interdisciplinary Cluster Environment (ALICE) provided by Leiden University.
 The views and opinions expressed here are solely those
of the authors and do not necessarily reflect those of the
funding institutions. Neither of the funding institutions
can be held responsible for them. 

\appendix
\section{The rank-one choice for two-body coefficients}\label{appendix:Discussionrank1}
Here we compare the Bell operator
$\hat I(\boldsymbol{\theta})$ in
Eq.~\eqref{eq:bell_operator_rankone}
with the Hamiltonian of the LMG model~\cite{lipkinValidityManybodyApproximation1965a}.
The key difference is the scaling of the coefficients with the system
size $N$. In the LMG model, the two-body coefficients are normalized by
$1/N$, while in the Bell operator they are independent
of $N$. As a result, the collective quadratic terms in
$\hat I(\boldsymbol{\theta})$ contribute at order $N^2$.

A robust large-$N$ violation requires an asymptotic ratio $\Delta=\beta_Q/\beta_C>1$. If the two-body coefficient matrix is full rank and positive definite, then any nonzero normalized collective correlator $\langle\mathcal S_k\rangle/N$ produces an $O(N^2)$ contribution. This leading-order contribution is the same in the quantum and classical optimizations, and therefore cannot generate an asymptotic separation. In the optimized quantum strategies considered here, the normalized collective correlators satisfy $\langle\mathcal S_k\rangle/N\to0$ for all measurements $k=0,\ldots,m-1$. For a full-rank positive-definite quadratic form, however, suppressing all these normalized correlators leaves no variables for optimizing the ratio.

A flat direction in the two-body correlators makes the separation possible. It removes the leading $O(N^2)$ contribution. The single-body terms and the squeezing corrections become the first nontrivial contributions to the separation between $\beta_Q$ and $\beta_C$. Since the measurements lie in the $x$–$z$ plane, a rank-one two-body coefficient matrix, as in Eq.~\eqref{eq:rankone_alpha} leaves one flat direction, characterized by $\sum_{k=0}^{m-1}\gamma_k\langle\mathcal S_k\rangle=0$. Along this direction, the leading $O(N^2)$ term vanishes, allowing spin squeezing to contribute at the first nontrivial order and making an asymptotic violation with $\Delta>1$ possible.

\section{Exact finite-size classical bounds by dynamic programming}
\label{appendix:dynamicprogramming}

We show a dynamic-programming method for computing
the finite-$N$ classical bound in the rank-one case. According to Eq.~\eqref{eq:I_Na_formgamma}, the classical bound is obtained
from the following integer program
\begin{align}
\begin{aligned}
\min_{\{N_{\boldsymbol{A}}\}}\quad&
\frac{1}{2}
\big(
\sum_{\boldsymbol{A}}
N_{\boldsymbol{A}}\,
\boldsymbol\gamma\cdot\boldsymbol{A}
\big)^2
+
\sum_{\boldsymbol{A}}
N_{\boldsymbol{A}}
\big[
\boldsymbol\alpha\cdot\boldsymbol{A}
-
\frac{1}{2}
\left(
\boldsymbol\gamma\cdot\boldsymbol{A}
\right)^2
\big]
\\
\mathrm{s.t.}\quad&
\sum_{\boldsymbol{A}}N_{\boldsymbol{A}}=N,
\qquad
N_{\boldsymbol{A}}\in\mathbb{Z}_{\ge0},
\end{aligned}
\label{eq:classical_bound_occupation_A}
\end{align}
where $\boldsymbol{A}$ is the strategy of a single party given in Eq.~\eqref{eq:Astrategy}.
Any negative sign in $\gamma_k$ can be absorbed into a sign flip of $A_k$, so without loss of generality we take all coefficients $\gamma_k$ to be positive and rescale them to integers for dynamic programming.

We simplify this optimization by grouping deterministic strategies with the
same value of
\begin{align}
u(\boldsymbol{A})
=
\boldsymbol\gamma\cdot\boldsymbol{A} .
\end{align}
Let 
$\mathcal U
=
\left\{
\boldsymbol\gamma\cdot\boldsymbol{A}:
\boldsymbol{A}\in\{\pm1\}^m
\right\}$
be the set of distinct rank-one contributions, and denote its cardinality by
\begin{align}
    M:=|\mathcal U|.
\end{align}
Since there are only $2^m$ deterministic single-site strategies, one has
$M\leq 2^m$. If $\boldsymbol{\gamma}$ has additional structure, this
number can be much smaller. For example, if all entries of
$\boldsymbol{\gamma}$ are equal, then $u(\boldsymbol A)$ only depends on
the number of $+1$ entries in $\boldsymbol A$, and hence $M=m+1$. For the paired structures in Eqs.~\eqref{eq:gammaeven} and \eqref{eq:sodd},
one obtains $M\leq 3^{m/2}$ for even $m$ and
$M\leq 2\cdot 3^{(m-1)/2}$ for odd $m$.

For each $u\in\mathcal U$, we keep only the strategy with the smallest linear
contribution and define
\begin{align}
w(u)
:=
\min_{\boldsymbol{A}:\,\boldsymbol\gamma\cdot\boldsymbol{A}=u}
\boldsymbol\alpha\cdot\boldsymbol{A} .
\end{align}
All other strategies with the same $u$ can be discarded, since the quadratic
part and the term $-u^2/2$ are fixed by $u$, while they have no smaller
linear contribution.

The minimization therefore reduces to an optimization over the occupation
numbers $N_u$ of the $M$ different $u$-values
\begin{align}
\begin{aligned}
\min_{\{N_u\}}\quad&
\frac{1}{2}
\left(
\sum_{u\in\mathcal U} N_u u
\right)^2
+
\sum_{u\in\mathcal U}
N_u
\left[
w(u)-\frac{u^2}{2}
\right]
\\
\mathrm{s.t.}\quad&
\sum_{u\in\mathcal U}N_u=N,
\qquad
N_u\in\mathbb{Z}_{\ge0}.
\end{aligned}
\label{eq:classical_bound_occupation_u}
\end{align}
Defining the shifted single-party cost
\begin{align}
    \widetilde w(u)
=
w(u)-\frac{u^2}{2},
\end{align}
the optimization can be written in terms of the accumulation
\begin{align}
q=\sum_{u\in\mathcal{U}}N_{u}u=\sum_{i=1}^{N}u_{i}.
\end{align}

With a fixed accumulated value $q$, the minimum cost after assigning $n$
parties can be computed by dynamic programming, closely related to the tropical-algebra approach used for translationally invariant Bell inequalities with few-body correlators \cite{huTropicalContractionTensor2026a,huCharacterizingTranslationinvariantBell2026a}. We initialize
\begin{align}
F_0(q)
&=
\begin{cases}
0, & q=0,\\
+\infty, & q\neq 0.
\end{cases}
\end{align}
For $n=1,\ldots,N$, we update
\begin{align}
F_{n}(q)=\min_{u_{n}\in\mathcal{U}}\left[F_{n-1}(q-u_{n})+\widetilde{w}(u_{n})+\frac{q^{2}-(q-u_{n})^{2}}{2}\right].
\label{eq:dp_Fn_update}
\end{align}
Then the classical bound is
\begin{align}
\beta_C
=
\min_q F_N(q).
\end{align}

If the coefficients $\gamma_k$ are positive integers, then every
$u\in\mathcal U$ is an integer. We denote
\begin{align}
G=\sum_{k=0}^{m-1}\gamma_k ,
\end{align}
so that $|u|\leq G$. After $n$ parties, the accumulated value satisfies
\begin{align}
-nG\leq q\leq nG.
\end{align}
Thus the number of possible choices of $q$ at step $n$ is at most $2nG+1$.
Since the recursion scans the $M$ allowed values of $u$, the total steps up to size $N$ is bounded by
\begin{align}
\sum_{n=0}^{N} M(2nG+1)
=
O(MGN^2),
\end{align}
which is the computational complexity reported in
Table~\ref{tab:DP_comparison}. The memory cost is $O(GN)$ if only two
consecutive dynamic programming layers are stored.

\section{Thermodynamic-limit classical bounds}\label{appendix:classicalbound}
Here we calculate the classical bounds in the thermodynamic limit $N\to\infty$.
We introduce the normalized occupation numbers
\begin{align}
p_{\boldsymbol{A}}:=\frac{N_{\boldsymbol{A}}}{N},
\qquad
p_{\boldsymbol{A}}\ge 0,
\qquad
\sum_{\boldsymbol{A}\in\{\pm1\}^m}p_{\boldsymbol{A}}=1 .
\end{align}
For finite $N$, these variables take values on a discrete grid. In the
thermodynamic limit, this grid becomes dense and the optimization can be
relaxed to an optimization over probability distributions
$p_{\boldsymbol A}$.

Throughout this limit, the coefficients $\boldsymbol{\alpha}$ and
$\boldsymbol{\gamma}$ are kept independent of $N$, and Eq.~\eqref{eq:I_Na_formgamma} can be written as
\begin{align}
I=\frac{N^{2}}{2}\big(\sum_{\boldsymbol{A}}p_{\boldsymbol{A}}\boldsymbol{\gamma}\cdot\boldsymbol{A}\big)^{2}+N\sum_{\boldsymbol{A}}p_{\boldsymbol{A}}\big[\boldsymbol{\alpha}\cdot\boldsymbol{A}-\frac{1}{2}(\boldsymbol{\gamma}\cdot\boldsymbol{A})^{2}\big].
\label{eq:I_pa_form}
\end{align}
The classical bound is obtained from the relaxed
optimization
\begin{align}
\beta_C
=
\min_{\substack{p_{\boldsymbol{A}}\ge 0,\\
\sum_{\boldsymbol{A}}p_{\boldsymbol{A}}=1}}
I.
\end{align}

The minimum in the large-$N$ limit is obtained by first minimizing the
non-negative $N^2$ term in Eq.~\eqref{eq:I_pa_form}. Hence the term ${N^{2}}\big(\sum_{\boldsymbol{A}}p_{\boldsymbol{A}}\boldsymbol{\gamma}\cdot\boldsymbol{A}\big)^{2}/2$ must vanish,
which gives
\begin{align}
\sum_{\boldsymbol A}
p_{\boldsymbol A}\,
\boldsymbol\gamma\cdot\boldsymbol A
=0 .
\end{align}
Under this constraint, the leading large-$N$ optimization becomes
\begin{align}
\begin{aligned}
\beta_{C}&=N\min_{p_{\boldsymbol{A}}}\sum_{\boldsymbol{A}}p_{\boldsymbol{A}}\big[\boldsymbol{\alpha}\cdot\boldsymbol{A}-\frac{1}{2}\left(\boldsymbol{\gamma}\cdot\boldsymbol{A}\right)^{2}\big],\\\mathrm{s.t.}\;&\sum_{\boldsymbol{A}}p_{\boldsymbol{A}}\boldsymbol{\gamma}\cdot\boldsymbol{A}=0,\;\sum_{\boldsymbol{A}}p_{\boldsymbol{A}}=1,\;p_{\boldsymbol{A}}\ge0.
\end{aligned}
\label{eq:betaC_pa_opt}
\end{align}

We regard $\boldsymbol{A}$ as a random variable with the probability distribution
$p_{\boldsymbol{A}}$, and the first moments and pair correlations are 
\begin{align}
    s_{k}&=\sum_{\boldsymbol{A}}p_{\boldsymbol{A}}A_{k},\quad z_{k,l}=\sum_{\boldsymbol{A}}p_{\boldsymbol{A}}A_{k}A_{l},
\end{align}
where $A_k$ is the entry of $\boldsymbol{A}$ (see Eq.~\eqref{eq:Astrategy}), and $z_{k,l}$ appears in $\sum_{\boldsymbol{A}}p_{\boldsymbol{A}}(\boldsymbol{\gamma}\cdot\boldsymbol{A})^{2}/2$. The linear functional can be written as 
\begin{align}
    I&=\frac{N^{2}}{2}\left(\sum_{k=0}^{m-1}\gamma_{k}s_{k}\right)^{2}+N(\sum_{k=0}^{m-1}\alpha_{k}s_{k}-\sum_{k,l=0}^{m-1}\frac{\gamma_{k}\gamma_{l}}{2}z_{k,l}).
\end{align}
We optimize over the pairwise correlations $z_{k,l}$ with fixed one-body moments $s_k$.

For the classical case, the deterministic strategy gives
$\langle A_k^{(i)}\rangle,\langle A_l^{(i)}\rangle\in\{\pm1\}$, and the
correlator can be written as
\begin{align}
\begin{aligned}
    \langle\mathcal Z_{k,l}\rangle
&=
N-\sum_{i=1}^{N}
\left|
\langle A_k^{(i)}\rangle
-
\langle A_l^{(i)}\rangle
\right|
\\&=
-N+\sum_{i=1}^{N}
\left|
\langle A_k^{(i)}\rangle
+
\langle A_l^{(i)}\rangle
\right|,
\end{aligned}
\end{align}
since $AB =1-|A-B|=-1+|A+B|$ when $A,B\in\{\pm1\}$.
The triangle inequality then gives the pairwise constraint
\begin{align}
-N+
\left|
\left\langle\mathcal S_k\right\rangle
+
\left\langle\mathcal S_l\right\rangle
\right|
\le
\left\langle\mathcal Z_{k,l}\right\rangle
\le
N-
\left|
\left\langle\mathcal S_k\right\rangle
-
\left\langle\mathcal S_l\right\rangle
\right|.
\label{ineq:Zkl}
\end{align}
The feasible range for $z_{k,l}$ follows from Eq.~\eqref{ineq:Zkl} as
\begin{align}
-1+|s_k+s_l|
\le z_{k,l}\le
1-|s_k-s_l|.
\end{align}
When all $\gamma_k > 0$, the coefficient of $z_{k,l}$ in the cost function is negative, so the minimum is achieved by taking the largest allowed value, i.e.,
\begin{align}
z_{k,l}=1-|s_k-s_l|.\label{eq:optimalzklwithsksl}
\end{align}

The following construction shows that these pairwise upper bounds can be
saturated simultaneously for all pairs $(k,l)$ by a strategy distribution $p_{\boldsymbol{A}}$. 
For arbitrary $s_k \in [-1,1]$, we define the random variable
\begin{align}
A_k(U) =
\begin{cases}
+1, & U \le s_k,\\
-1, & U > s_k,
\end{cases}
\end{align}
where $U$ is a random variable uniformly distributed on $[-1,1]$.
Each realization of $U$ specifies a deterministic strategy
\begin{align}
\boldsymbol{A}(U)
=
(A_0(U),\ldots,A_{m-1}(U))
\in\{\pm1\}^m,
\end{align}
which induces the probability distribution
\begin{align}
p_{\boldsymbol{A}}
=
\Pr\!\left(
\boldsymbol{A}(U)=\boldsymbol{A}
\right).
\end{align}
This construction satisfies
\begin{align}
\sum_{\boldsymbol{A}}
p_{\boldsymbol{A}}A_k
=
s_k .
\end{align}
Moreover, $A_k$ and $A_l$ differ only when $U$ lies between $s_k$ and $s_l$, which occurs with probability
\begin{align}
\Pr(A_k\neq A_l)
=
\frac{|s_k-s_l|}{2}.
\end{align}
It follows that
\begin{align}
\sum_{\boldsymbol{A}}
p_{\boldsymbol{A}}A_kA_l
=
1-2\Pr(A_k\neq A_l)
=
1-|s_k-s_l|.
\end{align}
Then we only need to optimize over the one-body variables $\boldsymbol s=(s_0,s_1,...,s_{m-1})$, which gives
\begin{align}
\beta_C
&=
\min_{\substack{s_k\in[-1,1],\\
\boldsymbol\gamma \cdot \boldsymbol s=0}}
N\big[
\sum_{k=0}^{m-1}\alpha_k s_k
-\frac12\sum_{k,l=0}^{m-1}\gamma_k\gamma_l\bigl(1-|s_k-s_l|\bigr)
\big].
\label{eq:betaCwithoutz}
\end{align}

To enforce the constraint $\boldsymbol{\gamma}\cdot \boldsymbol{s}
=0$, we adopt a paired structure in $\boldsymbol{\gamma}$ and $\boldsymbol{\alpha}$.
For even $m$, we parameterize the first $m/2$ elements with index $j$ starting from $1$ to $m/2$. These  vectors are explicitly given by Eq.~\eqref{eq:gammaeven}.
We then restrict to the antisymmetric sector 
\begin{align}
    \boldsymbol{s}=(s_{1},s_{2},\cdots,s_{m/2},-s_{1},-s_{2},\cdots,-s_{m/2}).
\end{align}
For odd $m$, these vectors are obtained similarly by prepending an additional element 
\begin{align}
\begin{aligned}
    \boldsymbol{\gamma}&=(\gamma_{0},\gamma_{1},\gamma_{2},\cdots,\gamma_{(m-1)/2},\gamma_{1},\gamma_{2},\cdots,\gamma_{(m-1)/2}),\\\boldsymbol{\alpha}&=(0,-\alpha_{1},-\alpha_{2},\cdots,-\alpha_{(m-1)/2},\alpha_{1},\alpha_{2},\cdots,\alpha_{(m-1)/2}),\\\boldsymbol{s}&=(0,s_{1},s_{2},\cdots,s_{(m-1)/2},-s_{1},-s_{2},\cdots,-s_{(m-1)/2}).
\end{aligned}\label{eq:sodd}
\end{align}
We assume $\alpha_j$ for $1\leq j\leq \left\lfloor m/2\right\rfloor$ are positive and ordered increasingly. Here $j$ labels the measurement pairs. With this convention, we restrict the minimization to the branch
$s_j\in[0,1]$ for $1\leq j\leq \left\lfloor m/2\right\rfloor$.

In the antisymmetric sector, the objective is piecewise linear in the variables $s_j$,
with linear regions determined by the ordering of the $s_j$'s. Since the
vertices of these regions have $s_j\in\{0,1\}$, the minimum can be attained
at a binary point, which gives
\begin{align}
\beta_{C}=\min_{s_{1},...,s_{m/2}\in\{0,1\}}-2N[\sum_{s_{j}=1}\alpha_{j}+(\sum_{s_{j}=0}\gamma_{j})^{2}],\label{eq:betaCeven}
\end{align}
for even $m$, and 
\begin{align}
\begin{aligned}
        \beta_{C}&=\min_{s_{1},...,s_{(m-1)/2}\in\{0,1\}}-2N[\sum_{s_{j}=1}\alpha_{j}+(\frac{1}{2}\gamma_{0}+\sum_{s_{j}=0}\gamma_{j})^{2}],\label{eq:betaCodd}
\end{aligned}
\end{align}
for odd $m$.
The even- and odd-$m$ results can be written in the unified form
\begin{align}
\beta_{C}=\frac{Nm^{2}}{2}\min_{s_{1},...,s_{\left\lfloor m/2\right\rfloor }\in\{0,1\}}g(\boldsymbol{s}),\label{eq:betaCcontiuum}
\end{align}
with
\begin{align}
g(\boldsymbol{s})&=\begin{cases}
{\displaystyle -\frac{4}{m^{2}}[\sum_{s_{j}=1}\alpha_{j}+(\sum_{s_{j}=0}\gamma_{j})^{2}],} & \mathrm{even}\;m\\
{\displaystyle -\frac{4}{m^{2}}[\sum_{s_{j}=1}\alpha_{j}+(\frac{\gamma_{0}}{2}+\sum_{s_{j}=0}\gamma_{j})^{2}],} & \mathrm{odd}\;m.
\end{cases}\label{eq:g_function}
\end{align}
We denote the minimum by $g^*=g(\boldsymbol{s}^*)$, which can be found by searching over the discrete choices of
$\boldsymbol{s}$. The normalized factor in Eq.~\eqref{eq:betaCcontiuum} gives a well-defined continuum limit of
$g(\boldsymbol{s})$ as $m\to\infty$, which is discussed in Appendix~\ref{sec:Nminftyresults}.

\section{Thermodynamic-limit quantum values}\label{appendix:quantumvalue}
Here we calculate the quantum values in the thermodynamic limit $N\to\infty$.
We rewrite the linear functional~\eqref{eq:I_rankone_collective} as 
\begin{align}
\begin{aligned}
    I&=\frac{1}{2}\left(\sum_{k=0}^{m-1}\gamma_{k}\left\langle \mathcal{S}_{k}\right\rangle \right)^{2}+\sum_{k=0}^{m-1}\alpha_{k}\left\langle \mathcal{S}_{k}\right\rangle -\sum_{k,l=0}^{m-1}\frac{\gamma_{k}\gamma_{l}}{2}\left\langle \mathcal{Z}_{k,l}\right\rangle \\&+\sum_{k,l=0}^{m-1}\frac{\gamma_{k}\gamma_{l}}{2}(\left\langle \mathcal{S}_{k}\mathcal{S}_{l}\right\rangle -\left\langle \mathcal{S}_{k}\right\rangle \left\langle \mathcal{S}_{l}\right\rangle ).\label{eq:Iquantum}
\end{aligned}
\end{align}
As in the classical case, we impose
$\sum_k\gamma_k s_k=0$, which removes the leading $O(N^2)$ term.
The covariance term in Eq.~\eqref{eq:Iquantum} is sublinear for a
spin-squeezed state along the direction
$\sum_k\gamma_k\mathcal S_k$.

We now use a semiclassical description in which the center of the spin-squeezed
state satisfies $S_x=(N\sin\phi) /2$ and $S_z =(N \cos \phi)/2 $, and we obtain $\langle\mathcal{S}_{k}\rangle=N\cos(\theta_{k}-\phi)$ and
$\langle\mathcal{Z}_{k,l}\rangle=N\cos(\theta_{k}-\theta_{l})$.
Similar to the classical case, we introduce
\begin{align}
s_{k}&:=\frac{\langle\mathcal{S}_{k}\rangle}{N}
=\cos(\theta_{k}-\phi),
\\
z_{k,l}&:=\frac{\langle\mathcal{Z}_{k,l}\rangle}{N}
=\cos(\theta_{k}-\theta_{l}).
\end{align}
The linear functional \eqref{eq:Iquantum} is simplified to
\begin{align}
\begin{aligned}
        I&=N^{2}\sum_{k,l=0}^{m-1}\frac{\gamma_{k}\gamma_{l}}{2}s_{k}s_{l}+N\big(\sum_{k=0}^{m-1}\alpha_{k}s_{k}-\sum_{k,l=0}^{m-1}\frac{\gamma_{k}\gamma_{l}}{2}z_{k,l}\big)\\&+
o(N),
\end{aligned}
\end{align}
Here the $o(N)$ term comes from the covariance term in
Eq.~\eqref{eq:Iquantum}.
The corresponding Gram matrix satisfies
\begin{align}
z_{k,k}=1,\qquad
\begin{pmatrix}
1 & \boldsymbol s^{\top}\\
\boldsymbol s & Z
\end{pmatrix}\succeq 0,
\end{align}
where $Z=(z_{k,l})_{k,l}$. The Gram matrix is generated by the unit vectors
$(\cos\phi,\sin\phi)$ and $(\cos\theta_k,\sin\theta_k)$. Since these vectors
lie in the $x$-$z$ plane, the Gram matrix has rank at most two.

For each pair $(k,l)$, the corresponding principal submatrix satisfies
\begin{align}
\begin{pmatrix}
1 & s_{k} & s_{l}\\
s_{k} & 1 & z_{k,l}\\
s_{l} & z_{k,l} & 1
\end{pmatrix}
\succeq0 .
\end{align}
This implies $s_ks_l-\sqrt{(1-s_k^2)(1-s_l^2)}
\le
z_{k,l}
\le
s_ks_l+\sqrt{(1-s_k^2)(1-s_l^2)} $.
Assume that all $\gamma_k\ge0$. Since the coefficient of $z_{k,l}$ in the
linear term is $-\gamma_k\gamma_l/2\le0$, the minimum is obtained by taking
the upper endpoint:
\begin{align}
z_{k,l}
=
s_ks_l+\sqrt{(1-s_k^2)(1-s_l^2)}.
\label{eq:z_plus_branch}
\end{align}
We now show that this choice of $z_{k,l}$ can be realized simultaneously
for all pairs. Without loss of generality, we choose the state direction along
the $x$ axis, i.e.,
\begin{align}
\phi=\pi/2 .
\end{align}
For each $k$, choose the measurement angle
\begin{align}
\theta_k=\arcsin(s_k).
\end{align}
For the two-body quantities, we obtain
\begin{align}
\begin{aligned}
   z_{k,l}= \cos(\theta_{k}-\theta_{l})
    &=s_{k}s_{l}+\sqrt{(1-s_{k}^{2})(1-s_{l}^{2})}.
\end{aligned}
\end{align}
Hence the choice in Eq.~\eqref{eq:z_plus_branch} is globally realizable.

Combining the constraint $\boldsymbol\gamma\cdot\boldsymbol s=0$ with the
globally realizable choice in Eq.~\eqref{eq:z_plus_branch}, the leading
thermodynamic-limit quantum value is
\begin{align}
\beta_{Q}
=\min_{\substack{s_{k}\in[-1,1],\\
\boldsymbol{\gamma}\cdot \boldsymbol{s}=0
}
}N[\sum_{k=0}^{m-1}\alpha_{k}s_{k}-\frac{1}{2}(\sum_{k=0}^{m-1}\gamma_{k}\sqrt{1-s_{k}^{2}})^{2}],\label{eq:betaQ}
\end{align}
where we use the fact that $\boldsymbol{\gamma}\cdot \boldsymbol{s}=\sum_{k=0}^{m-1}\gamma_{k}s_{k}=0$.
Using the same paired structure as in Eqs.~\eqref{eq:gammaeven} and
\eqref{eq:sodd}, Eq.~\eqref{eq:betaQ} can be written as
\begin{align}
\beta_{Q}=\frac{Nm^{2}}{2}\min_{s_{1},...,s_{\left\lfloor m/2\right\rfloor }\in[0,1]}f(\boldsymbol{s}),\label{eq:betaQwithfs}
\end{align}
with
\begin{align}
\begin{aligned}
    &f(\boldsymbol{s})=\\&\begin{cases}
    \displaystyle
\frac{-4}{m^{2}}[\sum_{j=1}^{m/2}\alpha_{j}s_{j}+(\sum_{j=1}^{m/2}\gamma_{j}\sqrt{1-s_{j}^{2}})^{2}], & \mathrm{even}\;m\\
\displaystyle
\frac{-4}{m^{2}}[\sum_{j=1}^{\frac{m-1}{2}}\alpha_{j}s_{j}+(\frac{\gamma_{0}}{2}+\sum_{j=1}^{\frac{m-1}{2}}\gamma_{j}\sqrt{1-s_{j}^{2}})^{2}], & \mathrm{odd}\;m.
\end{cases}
\end{aligned}\label{eq:f(s)simple}
\end{align}
We denote the minimum for the quantum value by 
$f^*=f(\boldsymbol{s}^*)$, where $\boldsymbol{s}^*=\{s_j^*\}_j$ is generally different from the
classical minimizer. The normalized factor in Eq.~\eqref{eq:betaQwithfs} gives a well-defined continuum limit of
$f(\boldsymbol{s})$ as $m\to\infty$, which is discussed in Appendix~\ref{sec:Nminftyresults}.

To determine the minimizer $\boldsymbol{s}^*$, we introduce  
\begin{align}
c=\begin{cases}
\displaystyle
\frac{2}{m}\sum_{j=1}^{m/2}\gamma_{j}\sqrt{1-s_{j}^{2}}, & \mathrm{even}\;m\\
\displaystyle
\frac{2}{m}\left(\frac{\gamma_{0}}{2}+\sum_{j=1}^{(m-1)/2}\gamma_{j}\sqrt{1-s_{j}^{2}}\right), & \mathrm{odd}\;m.
\end{cases}\label{eq:cwithxands}
\end{align}
For an interior minimum, each $s_j$ is chosen according to  
\begin{align}
\frac{\partial f}{\partial s_{j}}=-\frac{4}{m^{2}}[\alpha_{j}-\gamma_{j}\frac{mcs_{j}}{\sqrt{1-s_{j}^{2}}}]=0,
\end{align}
which gives 
\begin{align}
s_{j}^{*}=\frac{1}{\sqrt{1+m^{2}c^{2}\gamma_{j}^{2}/\alpha_{j}^{2}}}.
\label{eq:sj*}
\end{align}
Plugging Eq.~\eqref{eq:sj*} into Eq.~\eqref{eq:cwithxands}, we obtain a self-consistent equation to determine $c$ as 
\begin{align}
1=\begin{cases}
{\displaystyle \sum_{j=1}^{m/2}\frac{2\gamma_{j}^{2}}{\sqrt{\alpha_{j}^{2}+m^{2}c^{2}\gamma_{j}^{2}}},} & \mathrm{even}\;m\\
{\displaystyle \frac{\gamma_{0}}{mc}+\sum_{j=1}^{(m-1)/2}\frac{2\gamma_{j}^{2}}{\sqrt{\alpha_{j}^{2}+m^{2}c^{2}\gamma_{j}^{2}}},} & \mathrm{odd}\;m.
\end{cases}
\label{eq:selfconsistenteq}
\end{align}
Since Eq.~\eqref{eq:selfconsistenteq} does not generally admit a closed-form
solution for $c^*$, we first treat $c$ as a parameter. Substituting Eq.~\eqref{eq:sj*} into Eq.~\eqref{eq:f(s)simple}, we obtain
\begin{widetext}
\begin{align}
f(\boldsymbol{s}^{*},c)=\begin{cases}
\displaystyle
-\frac{4}{m^{2}}\sum_{j=1}^{m/2}\frac{\alpha_{j}^{2}}{\sqrt{\alpha_{j}^{2}+m^{2}c^{2}\gamma_{j}^{2}}}-\frac{4}{m^{2}}\left(\sum_{j=1}^{m/2}\gamma_{j}\frac{mc\gamma_{j}}{\sqrt{\alpha_{j}^{2}+m^{2}c^{2}\gamma_{j}^{2}}}\right)^{2}, & \mathrm{even}\;m\\
\displaystyle
-\frac{4}{m^{2}}\sum_{j=1}^{(m-1)/2}\frac{\alpha_{j}^{2}}{\sqrt{\alpha_{j}^{2}+m^{2}c^{2}\gamma_{j}^{2}}}-\frac{4}{m^{2}}\left(\frac{\gamma_{0}}{2}+\sum_{j=1}^{(m-1)/2}\gamma_{j}\frac{mc\gamma_{j}}{\sqrt{\alpha_{j}^{2}+m^{2}c^{2}\gamma_{j}^{2}}}\right)^{2}, & \mathrm{odd}\;m.
\end{cases}\label{eq:f(s*)simple}
\end{align}
\end{widetext}

\section{Infinite-$N$ optimization of quantum-to-classical ratios}\label{appendix:optimizingratioatinfiniteN}
Using the results of Appendices~\ref{appendix:classicalbound} and \ref{appendix:quantumvalue}, we optimize the ratio of the quantum value to the classical bound with odd and even number of measurements, respectively.

\subsection{General even $m$ measurements}
First, we optimize the ratio when the number of measurements $m$ is even. To obtain a tractable family of inequalities for optimization, we consider a sequence of vertices indexed by $p$ as  
\begin{align}
\begin{aligned}
    \boldsymbol{s}_{(p)}=&\bigl(\underbrace{0,\dots,0}_{p\ \text{zeros}},\underbrace{1,\dots,1}_{\frac{m}{2}-p\ \text{ones}},\underbrace{0,\dots,0}_{p\ \text{zeros}},\underbrace{-1,\dots,-1}_{\frac{m}{2}-p\ \text{minus ones}}\bigr),\\&0\leq p\leq m/2,
\end{aligned}
\end{align}
and impose the same classical bound
\begin{align}
g^*=-\frac{4}{m^{2}}[\left(\sum_{i=1}^{p}\gamma_{i}\right)^{2}+\sum_{j=p+1}^{m/2}\alpha_{j}],\;\mathrm{for\;any\;}0\leq p\leq m/2,\label{eq:g^*}
\end{align}
which gives $\left(\sum_{i=1}^{p}\gamma_{i}\right)^{2}+\sum_{j=p+1}^{m/2}\alpha_{j}=\left(\sum_{i=1}^{p-1}\gamma_{i}\right)^{2}+\sum_{j=p}^{m/2}\alpha_{j}$. Then we obtain the
recursive relation
\begin{align}
\alpha_{j}=\left(\sum_{i=1}^{j}\gamma_{i}\right)^{2}-\left(\sum_{i=1}^{j-1}\gamma_{i}\right)^{2}=2\gamma_{j}\left(\sum_{i=1}^{j}\gamma_{i}\right)-\gamma_{j}^{2}.\label{eq:alphakeven}
\end{align}
Since the classical bound is the same for all $p$, we represent it as $ g^{*}=-4\left(\sum_{i=1}^{m/2}\gamma_{i}\right)^{2}/m^{2}$.

Plugging Eq.~\eqref{eq:alphakeven} into Eq.~\eqref{eq:f(s)simple}, we rewrite the quantum value as
\begin{align}
f(\boldsymbol{s})=-\frac{4}{m^{2}}\{\sum_{j=1}^{m/2}[2\gamma_{j}(\sum_{k=1}^{j}\gamma_{k})-\gamma_{j}^{2}]s_{j}+[\sum_{j=1}^{m/2}\gamma_{j}\sqrt{1-s_{j}^{2}}]^{2}\}.
\end{align}
According to Eq.~\eqref{eq:sj*}, the optimal $s_j^*$ is given by 
\begin{align}
s_{j}^{*}=\left[1+\frac{m^{2}c^{*2}}{(\gamma_{j}+2\sum_{i=1}^{j-1}\gamma_{i})^{2}}\right]^{-1/2}.
\end{align}
Note that $c^*$ is determined by the self-consistent Eq.~\eqref{eq:selfconsistenteq}.

We further optimize the ratio $\Delta$ over the coefficients $\gamma_j$. We treat $c$ as a parameter and write the ratio according to Eqs.~\eqref{eq:f(s*)simple} and~\eqref{eq:g^*} as
\begin{align}
\begin{aligned}
    \Delta(\boldsymbol{\gamma},c)&=\sum_{j=1}^{m/2}\frac{1}{\left(\sum_{i=1}^{m/2}\gamma_{i}\right)^{2}}\frac{\gamma_{j}(\gamma_{j}+2\sum_{i=1}^{j-1}\gamma_{i})^{2}}{\sqrt{(\gamma_{j}+2\sum_{i=1}^{j-1}\gamma_{i})^{2}+m^{2}c^{2}}}\\&+\left(\sum_{j=1}^{m/2}\frac{\gamma_{j}}{\sum_{i=1}^{m/2}\gamma_{i}}\frac{mc}{\sqrt{(\gamma_{j}+2\sum_{i=1}^{j-1}\gamma_{i})^{2}+m^{2}c^{2}}}\right)^{2},
\end{aligned}
\end{align}
where $\boldsymbol{\gamma}$ and $c$ can be further optimized to maximize the ratio.
It is useful to introduce the cumulative variables
\begin{align}
\Gamma_{j}=\frac{\sum_{k=1}^{j}\gamma_{k}}{m/2},
\end{align}
and Eq.~\eqref{eq:alphakeven} becomes
\begin{align}
    \alpha_{k}&=\frac{m^{2}}{4}(\Gamma_{k}^{2}-\Gamma_{k-1}^{2}).\label{eq:alphakbyGammak}
\end{align}
With this change of variables, the quantity $\Delta$ depends on $\boldsymbol{\gamma}$ only through the sequence $\boldsymbol{\Gamma} = (\Gamma_1,\dots,\Gamma_{m/2})$, so we may regard it as a function of $\boldsymbol{\Gamma}$ and $c$, and write $\Delta(\boldsymbol{\Gamma},c)$. In terms of the $\Gamma_j$, the ratio can be rewritten as
\begin{align}
\begin{aligned}
    \Delta(\boldsymbol{\Gamma},c)&=\frac{1}{\Gamma_{m/2}^{2}}\Bigg[2\sum_{j=1}^{m/2}\frac{(\Gamma_{j}-\Gamma_{j-1})\left(\frac{\Gamma_{j}+\Gamma_{j-1}}{2}\right)^{2}}{\sqrt{\left(\frac{\Gamma_{j}+\Gamma_{j-1}}{2}\right)^{2}+c^{2}}}\\&+\left(\sum_{j=1}^{m/2}\frac{(\Gamma_{j}-\Gamma_{j-1})c}{\sqrt{\left(\frac{\Gamma_{j}+\Gamma_{j-1}}{2}\right)^{2}+c^{2}}}\right)^{2}\Bigg].
\end{aligned}\label{eq:DeltaXceven}
\end{align}
We fix $\Gamma_{m/2}=1$, and optimize $\boldsymbol{\Gamma}$ and $c$ to maximize the ratio. Below we give analytic solutions for $m=2,4,6$.

For $m=2$, we fix $\Gamma_1=1$, and the ratio simplifies to
\begin{align}
\Delta(c)&=\frac{1}{\sqrt{1+4c^{2}}}+\frac{4c^{2}}{1+4c^{2}}.
\end{align}
We find the optimal $c^*=\sqrt{3}/2$. The ratio and the corresponding Bell inequality are
\begin{align}
\Delta_{\infty,2}^{\mathrm{opt}}&=5/4,\;\boldsymbol{\gamma}^{*}=(1,1),\;\boldsymbol{\alpha}^{*}=(-1,1). \label{eq:ratioatm=2}
\end{align}

For $m=4$, we fix $\Gamma_2=1$, and the ratio simplifies to
\begin{align}
\begin{aligned}
    \Delta(\Gamma _{1},c)&=\frac{\Gamma_{1}^{3}}{\sqrt{\Gamma_{1}^{2}+4c^{2}}}+\frac{(1-\Gamma_{1})(1+\Gamma_{1})^{2}}{\sqrt{(1+\Gamma_{1})^{2}+4c^{2}}}\\&+4c^{2}\left(\frac{\Gamma_{1}}{\sqrt{\Gamma_{1}^{2}+4c^{2}}}+\frac{(1-\Gamma_{1})}{\sqrt{(1+\Gamma_{1})^{2}+4c^{2}}}\right)^{2}.
\end{aligned}\end{align}
The function can be maximized analytically with the optimal $\Gamma_1^*=15/34$, and $c^*=15^{3/2}/68$. The ratio and the corresponding Bell inequality are
\begin{align}
\begin{aligned}
\Delta_{\infty,4}^{\mathrm{opt}}&=353/272,\;\boldsymbol{\gamma}^{*}=(\frac{15}{17},\frac{19}{17},\frac{15}{17},\frac{19}{17}),\\\boldsymbol{\alpha}^{*}&=(-\frac{225}{289},-\frac{931}{289},\frac{225}{289},\frac{931}{289}).
\end{aligned}
\end{align}

For $m=6$, we fix $\Gamma_3=1$, and the ratio simplifies to
\begin{align}
    \begin{aligned}
\Delta(\Gamma_{1},\Gamma_{2},c)&=\frac{\Gamma_{1}^{3}}{\sqrt{\Gamma_{1}^{2}+4c^{2}}}+\frac{(\Gamma_{2}-\Gamma_{1})(\Gamma_{2}+\Gamma_{1})^{2}}{\sqrt{(\Gamma_{2}+\Gamma_{1})^{2}+4c^{2}}}\\&+\frac{(1-\Gamma_{2})(1+\Gamma_{2})^{2}}{\sqrt{(1+\Gamma_{2})^{2}+4c^{2}}}+4c^{2}\Big(\frac{\Gamma_{1}}{\sqrt{\Gamma_{1}^{2}+4c^{2}}}+\\&\frac{(\Gamma_{2}-\Gamma_{1})}{\sqrt{(\Gamma_{2}+\Gamma_{1})^{2}+4c^{2}}}+\frac{(1-\Gamma_{2})}{\sqrt{(1+\Gamma_{2})^{2}+4c^{2}}}\Big)^{2}.
    \end{aligned}
\end{align}
The function can be maximized analytically with the optimal $\Gamma_{1}^*=\frac{1225}{4251},\;\Gamma^*_{2}=\frac{2590}{4251}$, and $c^*=\frac{35^{5/2}}{8502}.$
The ratio and the corresponding Bell inequality are
\begin{align}
\begin{aligned}
\Delta_{\infty,6}^{\mathrm{opt}}&=66637/51012,\\\boldsymbol{\gamma}^{*}&=(\frac{1225}{1417},\frac{105}{109},\frac{1661}{1417},\frac{1225}{1417},\frac{105}{109},\frac{1661}{1417}),\\\boldsymbol{\alpha}^{*}&=(-\frac{1500625}{2007889},-\frac{3675}{1417},-\frac{11362901}{2007889},\\&\frac{1500625}{2007889},\frac{3675}{1417},\frac{11362901}{2007889}).
\end{aligned}
\end{align}

\subsection{General odd $m$ measurements}
We optimize the ratio for odd $m$. Following the same nested construction as in the even-$m$ case, the only difference is the presence of one unpaired entry, which is fixed along the sequence~\eqref{eq:sodd}.
We require
\begin{align}
\begin{aligned}
g^{*}&=-\frac{4}{m^{2}}\left[\left(\frac{\gamma_{0}}{2}+\sum_{i=1}^{j}\gamma_{i}\right)^{2}+\sum_{j=p+1}^{(m-1)/2}\alpha_{j}\right],\\&\;\mathrm{for\;any\;}0\leq p\leq(m-1)/2,
\end{aligned}
\end{align}
from which we solve $\alpha_j$ as 
\begin{align}
\alpha_{j}&=(\gamma_{0}+2\sum_{i=1}^{j}\gamma_{i})\gamma_{j}-\gamma_{j}^{2}.\label{eq:alphakodd}
\end{align}
Similarly to the case with even $m$, we represent the classical bound as $ g^{*}=-4\left(\gamma_{0}/2+\sum_{i=1}^{(m-1)/2}\gamma_{i}\right)^{2}/m^{2}$.
 For the quantum value, we simplify
 \begin{align}
\begin{aligned}
    f(\boldsymbol{s})&=-\frac{4}{m^{2}}\{\sum_{j=1}^{(m-1)/2}[\gamma_{0}\gamma_{j}+2\gamma_{j}(\sum_{k=1}^{j}\gamma_{k})-\gamma_{j}^{2}]s_{j}\\&+(\frac{\gamma_{0}}{2}+\sum_{j=1}^{(m-1)/2}\gamma_{j}\sqrt{1-s_{j}^{2}})^{2}\}.
\end{aligned}
 \end{align}
The optimal $s_j$ is given by 
\begin{align}
s_{j}^{*}=\left[1+\frac{m^{2}c^{*2}}{(\gamma_{0}+\gamma_{j}+2\sum_{i=1}^{j-1}\gamma_{i})^{2}}\right]^{-1/2}.
\end{align}
We treat $c$ as a parameter and obtain the ratio
\begin{widetext}
\begin{align}
\begin{aligned}
\Delta(\boldsymbol{\gamma},c)&=\sum_{j=1}^{(m-1)/2}\frac{1}{\left(\frac{\gamma_{0}}{2}+\sum_{i=1}^{(m-1)/2}\gamma_{i}\right)^{2}}\frac{\left[\gamma_{0}+\gamma_{j}+2\left(\sum_{i=1}^{j-1}\gamma_{i}\right)\right]^{2}\gamma_{j}}{\sqrt{(\gamma_{0}+\gamma_{j}+2\sum_{i=1}^{j-1}\gamma_{i})^{2}+m^{2}c^{2}}}\\&+\Bigg(\frac{\gamma_{0}}{2(\frac{\gamma_{0}}{2}+\sum_{i=1}^{(m-1)/2}\gamma_{i})}+\sum_{j=1}^{(m-1)/2}\frac{\gamma_{j}}{\frac{\gamma_{0}}{2}+\sum_{i=1}^{(m-1)/2}\gamma_{i}}\frac{mc}{\sqrt{(\gamma_{0}+\gamma_{j}+2\sum_{i=1}^{j-1}\gamma_{i})^{2}+m^{2}c^{2}}}\Bigg)^{2}.    
\end{aligned}
\end{align}
We introduce the cumulative variables $\Gamma_{j}=\frac{2}{m}(\frac{\gamma_{0}}{2}+\sum_{k=1}^{j}\gamma_{k})$ and $\Gamma_{0}=\frac{2}{m}\frac{\gamma_{0}}{2}$.
With these variables, Eq.~\eqref{eq:alphakodd} takes the same form as
Eq.~\eqref{eq:alphakbyGammak}.
Similarly to the case with even $m$, we rewrite the ratio into
\begin{align}
\begin{aligned}
\Delta(\boldsymbol{\Gamma},c)&=\frac{1}{\Gamma_{\frac{m-1}{2}}^{2}}\Big[2\sum_{j=1}^{\frac{m-1}{2}}\frac{\left(\frac{\Gamma_{j}+\Gamma_{j-1}}{2}\right)^{2}\bigl(\Gamma_{j}-\Gamma_{j-1}\bigr)}{\sqrt{\left(\frac{\Gamma_{j}+\Gamma_{j-1}}{2}\right)^{2}+c^{2}}}+\Big(\Gamma_{0}+\sum_{j=1}^{\frac{m-1}{2}}\frac{c\bigl(\Gamma_{j}-\Gamma_{j-1}\bigr)}{\sqrt{\left(\frac{\Gamma_{j}+\Gamma_{j-1}}{2}\right)^{2}+c^{2}}}\Big)^{2}\Big].
\label{eq:DeltaXcodd}
\end{aligned}
\end{align}
We fix $\Gamma_{(m-1)/2}=1$.
\end{widetext}

For $m=3$, we fix $\Gamma_1=1$, and the ratio simplifies to
\begin{align}
\begin{aligned}
\Delta(\Gamma_{0},c)&=\frac{(1+\Gamma_{0})^{2}(1-\Gamma_{0})}{\sqrt{(1+\Gamma_{0})^{2}+4c^{2}}}+\Big(\Gamma_{0}+\frac{c(1-\Gamma_{0})}{\sqrt{(\frac{1+\Gamma_{0}}{2})^{2}+c^{2}}}\Big)^{2}.
\end{aligned}
\end{align}
We find the optimal $\Gamma^*_0=2/7$ and $ c^*=6/7$.
The ratio and the corresponding Bell inequality are
\begin{align}
\begin{aligned}
\Delta_{\infty,3}^{\mathrm{opt}}&=9/7,\quad\boldsymbol{\gamma}^{*}=(\frac{6}{7},\frac{15}{14},\frac{15}{14}),\\\boldsymbol{\alpha}^{*}&=(0,-\frac{405}{196},\frac{405}{196}).
\end{aligned}
\label{eq:ratioatm=3}
\end{align}

For $m=5$, we fix $\Gamma_2=1$, and the ratio simplifies to
\begin{align}
\begin{aligned}
    \Delta(\Gamma_{0},\Gamma_{1},c)&=\frac{(\Gamma_{1}+\Gamma_{0})^{2}(\Gamma_{1}-\Gamma_{0})}{\sqrt{(\Gamma_{1}+\Gamma_{0})^{2}+4c^{2}}}+\frac{(1+\Gamma_{1})^{2}(1-\Gamma_{1})}{\sqrt{(1+\Gamma_{1})^{2}+4c^{2}}}\\&+\Bigg(\Gamma_{0}+\frac{c(\Gamma_{1}-\Gamma_{0})}{\sqrt{\left(\frac{\Gamma_{1}+\Gamma_{0}}{2}\right)^{2}+c^{2}}}+\frac{c(1-\Gamma_{1})}{\sqrt{\left(\frac{1+\Gamma_{1}}{2}\right)^{2}+c^{2}}}\Bigg)^{2}.
\end{aligned}
\end{align}
We find the optimal
 $\Gamma^*_0=\frac{36}{211}$, $\Gamma^*_1=\frac{114}{211}$, and $c^*=\frac{180}{211}$.
The ratio and the corresponding Bell inequality are 
\begin{align}
\begin{aligned}
\Delta_{\infty,5}^{\mathrm{opt}}&={275}/{211},\quad\boldsymbol{\gamma}^{*}=(\frac{180}{211},\frac{195}{211},\frac{485}{422},\frac{195}{211},\frac{485}{422}),\\\boldsymbol{\alpha}^{*}&=(0,-\frac{73125}{44521},-\frac{788125}{178084},\frac{73125}{44521},\frac{788125}{178084}).
\end{aligned}
\end{align}

\subsection{Uniform weights $\gamma_j=1$}
Here we consider the case with fixed $\gamma_j=1$ and
$\alpha_j$ determined by Eqs. \eqref{eq:alphakeven} and \eqref{eq:alphakodd}, corresponding to the PI Bell inequalities given in
Ref.~\cite{wagnerBellCorrelationsManyBody2017}. The quantum-to-classical ratios of these Bell inequalities are given in Table~\ref{tab:infiniteN-different-m}. From
Eqs.~\eqref{eq:alphakeven} and~\eqref{eq:alphakodd}, we obtain
$\alpha_j=2j-1$ for even $m$ and $\alpha_j=2j$ for odd $m$.
In this case, the ratio simplifies to
\begin{widetext}
\begin{align}
\Delta(\boldsymbol{1},c)
&=\begin{cases}
\displaystyle
\sum_{j=1}^{m/2}\frac{4}{m^{2}}\frac{(2j-1)^{2}}{\sqrt{(2j-1)^{2}+m^{2}c^{2}}}
+\left(\sum_{j=1}^{m/2}\frac{2c}{\sqrt{(2j-1)^{2}+m^{2}c^{2}}}\right)^{2},
& \text{even } m,\\[1.2em]
\displaystyle
\sum_{j=1}^{(m-1)/2}\frac{4}{m^{2}}\frac{4j^{2}}{\sqrt{4j^{2}+m^{2}c^{2}}}
+\left(\frac{1}{m}+\sum_{j=1}^{(m-1)/2}\frac{2c}{\sqrt{4j^{2}+m^{2}c^{2}}}\right)^{2},
& \text{odd } m.
\end{cases}
\label{eq:Delta_uniform_c}
\end{align}
In Table~\ref{tab:infiniteN-different-m}, the values of the ratio for
finite $m$ and fixed $\boldsymbol{\gamma}=\boldsymbol{1}$ are obtained
from Eq.~\eqref{eq:Delta_uniform_c} by numerically maximizing
$\Delta(\boldsymbol{1},c)$ over $c$. In the large-$m$ limit, the sums in
Eq.~\eqref{eq:Delta_uniform_c} can be replaced by integrals, giving the same
continuum expression in Eq.~\eqref{eq:Deltaintegralresult} with
$\Gamma_{\mathrm f}=1$.

\subsection{Number of measurement $m\to\infty$}\label{sec:Nminftyresults}
In the limit $m\to\infty$, we represent the discrete sequences
$\{\alpha_j\}$, $\{\gamma_j\}$, and $\{s_j\}$ by smooth functions
$\alpha(u)$, $\gamma(u)$, and $s(u)$ on $[0,1]$.
We define them by
$\alpha_j=m\alpha(2j/m)$, $\gamma_j=\gamma(2j/m)$, and
$s_j=s(2j/m)$. $\Gamma_j$ becomes
\begin{align}
    \Gamma_{j}=\frac{\sum_{k=1}^{j}\gamma_{k}}{m/2}=\frac{\sum_{k=1}^{j}\gamma(2k/m)}{m/2}\approx\int_{0}^{2j/m}\gamma(u)\,du.
\end{align}
This scaling is chosen so that the summations have a finite continuum limit.

For the paired construction, we take $0\le s(u)\le 1$, and simplify the classical bound as a functional of $s(u)$:
\begin{align}
g[s(u)]
&= -\int_{0}^{1} 2\alpha(u)s(u)\,du
   -\iint_{0}^{1}du\,du^{\prime}\,\gamma(u)\gamma(u^{\prime})
     \bigl[1-\max\bigl(s(u),s(u^{\prime})\bigr)\bigr].
\end{align}
Using the fact that the optimal $s(u)$ takes only the values $0$ or $1$, we write $E=\{u\in[0,1]:\,s(u)=1\}$ and obtain
\begin{align}
g[s(u)]
= -\int_{E}2\alpha(u)\,du
  -\Bigl(\int_{E^{c}}\gamma(u)\,du\Bigr)^{2},
\end{align}
where $E^{c}$ is the complement of $E$.

The quantum value is related to the functional
\begin{align}
f[s(u)]
&= -\int_{0}^{1}2\alpha(u)s(u)\,du
   -\left(\int_{0}^{1}\gamma(u)\sqrt{1-s(u)^{2}}\,du\right)^{2},
\end{align}
where $s(u)$ satisfies the constraint $0\leq s(u)\leq 1$, and we denote
\begin{align}
    c &= \int_{0}^{1}\gamma(u)\sqrt{1-s(u)^{2}}\,du.
\end{align}
The optimal $s^{*}(u)$ is
\begin{align}
s^{*}(u)
= \frac{1}{\sqrt{1+c^{2}\gamma(u)^{2}/\alpha(u)^{2}}},
\end{align}
and the self-consistency condition for $c$ becomes
\begin{align}
    \int_{0}^{1}\frac{\gamma(u)^{2}}{\sqrt{\alpha(u)^{2}+c^{2}\gamma(u)^{2}}}\,du = 1.
\end{align}

We maximize the ratio $\Delta$ by optimizing the functional forms of $\alpha(u)$ and $\gamma(u)$ in the following. The optimized choice of $\alpha(u)$ follows from Eqs.~\eqref{eq:alphakeven} and~\eqref{eq:alphakodd} as
\begin{align}
    \alpha(u) &= \gamma(u)\int_{0}^{u}\gamma(v)\,dv.
    \label{eq:alphacontinuouslimit}
\end{align}
The corresponding optimal classical bound is
\begin{align}
    g^{*} = -\left[\int_{0}^{1}\gamma(u)\,du\right]^{2}.
\end{align}
With $\alpha(u)$ given by Eq.~\eqref{eq:alphacontinuouslimit}, the quantum value is related to
\begin{align}
f[s(u),\gamma(u)]
&= -2\int_{0}^{1}du\int_{0}^{u}dv\,\gamma(u)\gamma(v)s(u)
   -\left(\int_{0}^{1}\gamma(u)\sqrt{1-s(u)^{2}}\,du\right)^{2},
\end{align}
by choosing
\begin{align}
s^{*}(u)
&=\frac{1}{\sqrt{1+c^{*2}/\left[\int_{0}^{u}\gamma(v)\,dv\right]^{2}}}.
\end{align}
Substituting the optimal $s^{*}(u)$ into $f[s(u)]$, we obtain
\begin{align}
f^{*}[\Gamma(u)]
&=-2\int_{0}^{1}du\,
   \frac{\Gamma(u)^{2}\Gamma'(u)}
        {\sqrt{\Gamma(u)^{2}+c^{*2}}}
   -\left(
      \int_{0}^{1}du\,\frac{c^{*}\Gamma'(u)}
        {\sqrt{\Gamma(u)^{2}+c^{*2}}}
    \right)^{2}, \label{eq:fstar-Gamma-u}
\end{align}
where we use the cumulative function 
\begin{align}
    \Gamma(u)=\int_{0}^{u}\gamma(v)\,dv, \qquad \Gamma'(u)=\gamma(u).
\end{align}

Since $\gamma(u)>0$, $\Gamma(u)$ is monotonic and the change of variables
$\Gamma=\Gamma(u)$ is valid. We denote $\Gamma_{\mathrm f}:=\Gamma(1)$. Then, Eq.~\eqref{eq:fstar-Gamma-u} is equivalently
\begin{align}
f^{*}[\Gamma]
&=-2\int_{0}^{\Gamma_{\mathrm f}}d\Gamma\,
   \frac{\Gamma^{2}}
        {\sqrt{\Gamma^{2}+c^{*2}}}
   -\left(
      \int_{0}^{\Gamma_{\mathrm f}}d\Gamma\,\frac{c^{*}}
        {\sqrt{\Gamma^{2}+c^{*2}}}
    \right)^{2}. \label{eq:fstar-Gamma}
\end{align}
In the continuum limit $m\rightarrow\infty$, the discrete expressions~\eqref{eq:DeltaXceven} and~\eqref{eq:DeltaXcodd} for the ratio $\Delta$ also reduce to an integral that depends only on the final cumulative value $\Gamma_{m/2}=\Gamma_{\mathrm f}$. The ratio is explicitly
\begin{align}
    \Delta(\Gamma_{\mathrm f},c)
    &=\frac{1}{\Gamma_{\mathrm f}^{2}}\left[
        2\int_{0}^{\Gamma_{\mathrm f}}\frac{\Gamma^{2}}{\sqrt{\Gamma^{2}+c^{2}}}\,d\Gamma
        +\left(\int_{0}^{\Gamma_{\mathrm f}}\frac{c}{\sqrt{\Gamma^{2}+c^{2}}}\,d\Gamma\right)^{2}
    \right].    \label{eq:Deltaintegralresult}
\end{align}
Thus, in the double limit $N\to\infty$ and $m\to \infty$, both the optimal quantum value and the classical bound depend on $\gamma(u)$ only through the final cumulative parameter $\Gamma_{\mathrm f}$, and the ratio $\Delta$ is fully characterized by the one-dimensional integral in Eq.~\eqref{eq:Deltaintegralresult}.
We fix $\Gamma_{m/2}=1$ and complete the integral to obtain 
\begin{align}
\Delta_{\infty,\infty}(c)
=
\sqrt{1+c^{2}}
+
c^{2}
\left[
\operatorname{arccoth}^{2}\!\left(\sqrt{1+c^{2}}\right)
-
\operatorname{arccoth}\!\left(\sqrt{1+c^{2}}\right)
\right].
\end{align}
The optimized ratio is
\begin{align}
    \Delta_{\infty,\infty}=\coth (1),
\end{align}
which is reached by 
\begin{align}
    c^*=1/\sinh(1).
\end{align}

\section{Certification of the asymptotic quantum value}\label{appendix:certification_approach}
Besides the variational approach described in Sec.~\ref{sec:Variational approach}, we certify the quantum value using a sum-of-squares decomposition of the shifted linear functional
\begin{align}
\tilde I:=I-\beta_Q
\end{align} 
If $\tilde I$ admits a sum-of-squares decomposition, then
$\tilde I\succeq0$ and the corresponding value provides a certified lower
bound on the quantum value. We formulate this certification as a semidefinite program.
\subsection{Sum-of-squares formulation}

Using Eq.~\eqref{eq:Iwithuptotwobody}, we obtain
\begin{align}
\tilde I
=
-\frac{N}{2}\sum_{k=0}^{m-1}\alpha_{k,k}
-\beta_Q
+
\sum_{k=0}^{m-1}\alpha_k\mathcal S_k
+
\frac12\sum_{k=0}^{m-1}\alpha_{k,k}\mathcal S_k^2
+
\frac12
\sum_{\substack{k,l=0\\k\neq l}}^{m-1}
\alpha_{k,l}
\left(
\mathcal S_k\mathcal S_l-\mathcal Z_{k,l}
\right).
\label{eq:IminusbetaQdefinetildeB}
\end{align}
Here we use $\alpha_{k,l}$ for the general two-body coefficients.
We seek a sum-of-squares decomposition of the form
\begin{align}
\tilde I
=
\sum_{p=0}^{m-1}
\left(
\mu^{(p)}
+
\sum_{k=0}^{m-1}
\mu_k^{(p)}\mathcal S_k
\right)^2
+
\sum_{q\in\mathrm{pairs}}
\sum_{k,l\in q}
\nu_k^{(q)}
\nu_l^{(q)}
\left(
N\mathcal Z_{k,l}
-
\mathcal S_k\mathcal S_l
\right),
\label{eq:SOSoftildeB}
\end{align}
where $q=\{k,l\}$ denotes an unordered pair of measurement settings.
The first term is manifestly positive semidefinite. 
To rewrite the second term
as a sum of squares, we define
\begin{align}
B^{(i)}_{k,l}
=
\nu_k^{(k,l)}A_k^{(i)}
+
\nu_l^{(k,l)}A_l^{(i)},
\end{align}
and rewrite 
\begin{align}
\begin{aligned}
    \sum_{p,q\in\{k,l\}}\nu_{p}^{(k,l)}\nu_{q}^{(k,l)}\left(N\mathcal{Z}_{pq}-\mathcal{S}_{p}\mathcal{S}_{q}\right)&=N\sum_{i=1}^{N}\left(B_{k,l}^{(i)}\right)^{2}-\left(\sum_{i=1}^{N}B_{k,l}^{(i)}\right)^{2}\\&=N\sum_{a=1}^{N-1}\frac{\left(\sum_{i=1}^{a}B_{k,l}^{(i)}-aB_{k,l}^{(a+1)}\right)^{2}}{a(a+1)}.
\end{aligned}\label{eq:varianceSOSidentity}
\end{align}
Thus each pair contribution in Eq.~\eqref{eq:SOSoftildeB} is positive
semidefinite.

Expanding Eq.~\eqref{eq:SOSoftildeB}, we obtain
\begin{align}
\begin{aligned}
\tilde I
&=
\sum_{p=0}^{m-1}
\left(\mu^{(p)}\right)^2
+
2\sum_{k=0}^{m-1}
\left(
\sum_{p=0}^{m-1}
\mu^{(p)}\mu_k^{(p)}
\right)\mathcal S_k
+
\sum_{k,l=0}^{m-1}
\left(
\sum_{p=0}^{m-1}
\mu_k^{(p)}\mu_l^{(p)}
-
E_{k,l}
\right)
\mathcal S_k\mathcal S_l
+
N\sum_{k,l=0}^{m-1}
E_{k,l}\mathcal Z_{k,l}.
\end{aligned}
\label{eq:expandedSOS}
\end{align}
Here the pair contribution is collected into a symmetric matrix
$E=(E_{k,l})$ defined as follows. For each unordered pair
$q=\{k,l\}$, define a column vector $\boldsymbol{\nu}^{(q)}\in\mathbb R^m$
whose only nonzero components are $\nu_k^{(q)}$ and $\nu_l^{(q)}$.
Then the symmetric matrix is positive semidefinite
\begin{align}
E
=
\sum_{q\in\mathrm{pairs}}
\boldsymbol{\nu}^{(q)}
\left(\boldsymbol{\nu}^{(q)}\right)^T \succeq 0
\end{align}
Equivalently, in components,
\begin{align}
    E_{k,l}
=
\begin{cases}
\displaystyle
\nu_k^{(\{k,l\})}
\nu_l^{(\{k,l\})},
&
k\neq l,
\\[0.8em]
\displaystyle
\sum_{\substack{q\in\mathrm{pairs}\\ k\in q}}
\left(
\nu_k^{(q)}
\right)^2,
&
k=l .
\end{cases}
\label{eq:Edefinition}
\end{align}
The pair term can thus be written compactly as
\begin{align}
    \sum_{q\in\mathrm{pairs}}
\sum_{k,l\in q}
\nu_k^{(q)}
\nu_l^{(q)}
\left(
N\mathcal Z_{k,l}
-
\mathcal S_k\mathcal S_l
\right)=\sum_{k,l=0}^{m-1}
E_{k,l}
\left(
N\mathcal Z_{k,l}
-
\mathcal S_k\mathcal S_l
\right).
\end{align}
Since $\mathcal Z_{k,k}=N$, the diagonal part contributes only to the
constant term and to the coefficients of $\mathcal S_k^2$.

\subsection{Semidefinite Programming Reduction}

Expanding the first square term in
Eq.~\eqref{eq:SOSoftildeB} gives
\begin{align}
&\sum_{p=0}^{m-1}
\left(
\mu^{(p)}
+
\sum_{k=0}^{m-1}
\mu_k^{(p)}\mathcal S_k
\right)^2
=
\sum_{p=0}^{m-1}
(\mu^{(p)})^2
+
2\sum_{k=0}^{m-1}
\left(
\sum_{p=0}^{m-1}
\mu^{(p)}\mu_k^{(p)}
\right)\mathcal S_k
+
\sum_{k,l=0}^{m-1}
\left(
\sum_{p=0}^{m-1}
\mu_k^{(p)}\mu_l^{(p)}
\right)
\mathcal S_k\mathcal S_l .
\label{eq:firstSOSexpansion}
\end{align}
Since the coefficients are generated by quadratic forms, they define a positive semidefinite matrix,
\begin{align}
\begin{pmatrix}
\sum_p(\mu^{(p)})^2
&
\sum_p\mu^{(p)}\mu_0^{(p)}
&
\cdots
&
\sum_p\mu^{(p)}\mu_{m-1}^{(p)}
\\
\sum_p\mu^{(p)}\mu_0^{(p)}
&
\sum_p(\mu_0^{(p)})^2
&
\cdots
&
\sum_p\mu_0^{(p)}\mu_{m-1}^{(p)}
\\
\vdots
&
\vdots
&
\ddots
&
\vdots
\\
\sum_p\mu^{(p)}\mu_{m-1}^{(p)}
&
\sum_p\mu_{m-1}^{(p)}\mu_0^{(p)}
&
\cdots
&
\sum_p(\mu_{m-1}^{(p)})^2
\end{pmatrix}
\succeq0 .
\label{eq:firstGramPSD}
\end{align}

Comparing
Eq.~\eqref{eq:expandedSOS}
with
Eq.~\eqref{eq:IminusbetaQdefinetildeB},
we obtain
\begin{align}
\sum_{p=0}^{m-1}
(\mu^{(p)})^2
+
N^2
\sum_{k=0}^{m-1}
E_{k,k}
&=
-\frac{N}{2}
\sum_{k=0}^{m-1}
\alpha_{k,k}
-\beta_Q ,
\\
2
\sum_{p=0}^{m-1}
\mu^{(p)}
\mu_k^{(p)}
&=
\alpha_k ,
\\
\sum_{p=0}^{m-1}
(\mu_k^{(p)})^2
-
E_{k,k}
&=
\frac12\alpha_{k,k},
\\
\sum_{p=0}^{m-1}
\mu_k^{(p)}
\mu_l^{(p)}
-
E_{k,l}
&=
\frac12\alpha_{k,l},
\qquad
k\neq l ,
\\
NE_{k,l}
&=
-\frac12\alpha_{k,l},
\qquad
k\neq l .
\label{eq:Eoffdiag}
\end{align}
The last equation fixes the off-diagonal entries,
\begin{align}
E_{k,l}
=
-\frac{\alpha_{k,l}}{2N},
\qquad
k\neq l ,
\label{eq:Eoffdiagfixed}
\end{align}
while the diagonal entries $E_{k,k}$ remain free variables constrained by $E\succeq0$.

Substituting these relations into
Eq.~\eqref{eq:firstGramPSD},
we obtain the semidefinite constraint
\begin{align}
&
\begin{pmatrix}
-\beta_Q
-\frac{N}{2}\sum_{j=0}^{m-1}\alpha_{j,j}
-
N^2\sum_{j=0}^{m-1}E_{j,j}
&
\frac{\alpha_0}{2}
&
\cdots
&
\frac{\alpha_{m-1}}{2}
\\
\frac{\alpha_0}{2}
&
\frac{\alpha_{0,0}}{2}+E_{0,0}
&
\cdots
&
\frac{\alpha_{0,m-1}}{2}
-\frac{\alpha_{0,m-1}}{2N}
\\
\vdots
&
\vdots
&
\ddots
&
\vdots
\\
\frac{\alpha_{m-1}}{2}
&
\frac{\alpha_{0,m-1}}{2}
-\frac{\alpha_{0,m-1}}{2N}
&
\cdots
&
\frac{\alpha_{m-1,m-1}}{2}
+
E_{m-1,m-1}
\end{pmatrix}
\succeq0 .
\label{eq:mainSDPmatrix}
\end{align}
\end{widetext}

Meanwhile, $E\succeq0$ is explicitly
\begin{align}
\begin{pmatrix}
E_{0,0}
&
-\frac{\alpha_{0,1}}{2N}
&
\cdots
&
-\frac{\alpha_{0,m-1}}{2N}
\\
-\frac{\alpha_{0,1}}{2N}
&
E_{1,1}
&
\cdots
&
-\frac{\alpha_{1,m-1}}{2N}
\\
\vdots
&
\vdots
&
\ddots
&
\vdots
\\
-\frac{\alpha_{0,m-1}}{2N}
&
-\frac{\alpha_{1,m-1}}{2N}
&
\cdots
&
E_{m-1,m-1}
\end{pmatrix}
\succeq0 .
\label{eq:ESDPmatrix}
\end{align}

Therefore, the semidefinite programming (SDP) relaxation is
\begin{align}
\beta_Q^{\mathrm{1-level}}
=
\max_{\beta_Q,\{E_{k,k}\}}
\beta_Q
\end{align}
subject to the semidefinite constraints
Eqs.~\eqref{eq:mainSDPmatrix}
and~\eqref{eq:ESDPmatrix}.
For fixed Bell coefficients and system size $N$, this SDP provides a
certified lower bound on the quantum value. For the PI Bell inequalities
given by Eqs.~\eqref{eq:bellm=2},~\eqref{eq:bellm=3},~\eqref{eq:bellm=4},~\eqref{eq:bellm=5}, and~\eqref{eq:bellm=6}, the certified bounds
obtained from the SDP agree with the corresponding variational
results under the numerical error of the SDP solver.

\section{Certification of the LMG-like ground-state energy of Eq.~\eqref{eq:I_spin_squeezed}}\label{appendix:certifyLMGenergy}

Here we give the certified lower bound for the ground-state energy for the Hamiltonian of the LMG model.
The corresponding optimization problem is 
\begin{align}
\begin{aligned}
    \min_{(\boldsymbol{S},\Gamma)\in\mathcal{R}}\Big[J+J_{x}S_{x}+J_{z}S_{z}+\frac{1}{2}(\gamma_{x}S_{x}+\gamma_{z}S_{z})^{2}\\+\frac{1}{2}\bigl(\gamma_{x}^{2}\Gamma_{x,x}+2\gamma_{x}\gamma_{z}\Gamma_{x,z}+\gamma_{z}^{2}\Gamma_{z,z}\bigr)\Big],
\end{aligned}
\label{eq:moment_optimization_general}
\end{align}
where $\mathcal{R}$ denotes the set of allowed first and second moments for the spin-$N/2$ irreducible representation, and $S_\mu$ and $\Gamma_{\mu,\nu}$ are given by Eqs. \eqref{eq:Smumean} and \eqref{eq:Gammamunucovariance}, respectively. 
Any feasible point in $\mathcal R$ must in particular satisfy
\begin{align}
\begin{aligned}
S_x^2+S_y^2+S_z^2+\Gamma_{x,x}+\Gamma_{y,y}+\Gamma_{z,z}&= S(S+1),\\
    S_x^2 +S_y^2 +S_z^2 &\leq S^2,
\end{aligned}\label{eq:certifiedconstraint1}
\end{align}
with $S=N/2$.
The spin components and covariance entries are quantum moment variables and
cannot be chosen independently.  They must satisfy the uncertainty relation according to the commutation
relations
$[\hat S_\mu,\hat S_\nu]=i\epsilon_{\mu\nu\rho}\hat S_\rho$.  Equivalently,
the covariance matrix of arbitrary complex linear combinations of spin
fluctuations must be positive semidefinite, which gives
\begin{align}
\begin{pmatrix}
\Gamma_{x,x} & \Gamma_{x,y}+\frac{i}{2}S_z & \Gamma_{x,z}-\frac{i}{2}S_y\\
\Gamma_{x,y}-\frac{i}{2}S_z & \Gamma_{y,y} & \Gamma_{y,z}+\frac{i}{2}S_x\\
\Gamma_{x,z}+\frac{i}{2}S_y & \Gamma_{y,z}-\frac{i}{2}S_x & \Gamma_{z,z}
\end{pmatrix}
&\succeq 0 .
\label{eq:certifiedconstraint2}
\end{align}

For fixed first moments $S_x$ and $S_z$, we optimize the covariance contribution
\begin{align}
\Gamma_{\gamma,\gamma}:=\frac{\gamma_{x}^{2}\Gamma_{x,x}+2\gamma_{x}\gamma_{z}\Gamma_{x,z}+\gamma_{z}^{2}\Gamma_{z,z}}{\gamma_{x}^{2}+\gamma_{z}^{2}}.
\end{align}
Since the objective involves only the $x$ and $z$ components, the mean spin can be chosen in the $x$-$z$ plane. We therefore set $S_y=0$.
Optimizing $\Gamma_{\gamma,\gamma}$ under the constraint \eqref{eq:certifiedconstraint2} gives
\begin{align}
\begin{aligned}
\Gamma_{\gamma,\gamma}^{\min}&=\frac{(\gamma_{x}S_{z}-\gamma_{z}S_{x})^{2}}{2(\gamma_{x}^{2}+\gamma_{z}^{2})(S_{x}^{2}+S_{z}^{2})}\Big[S\left(S+1\right)-(S_{x}^{2}+S_{z}^{2})\\&-\sqrt{[S\left(S+1\right)-(S_{x}^{2}+S_{z}^{2})]^{2}-(S_{x}^{2}+S_{z}^{2})}\Big].
\end{aligned}
\label{eq:Gamma_gammagamma_min_SxSz}
\end{align}
The case $S_x=S_z=0$ can be treated separately and is not relevant for the
minimum considered here.

After the covariance optimization, the remaining optimization over the mean-field terms $S_x$ and $S_z$ becomes
\begin{align}
\min_{S_x,S_z} f(S_x,S_z),
\qquad
S_x^2+S_z^2\le S^2,
\end{align}
where
\begin{align}
\begin{aligned}
f(S_{x},S_{z})&=J+J_{x}S_{x}+J_{z}S_{z}+\frac{1}{2}(\gamma_{x}S_{x}+\gamma_{z}S_{z})^{2}\\&+\frac{(\gamma_{x}S_{z}-\gamma_{z}S_{x})^{2}}{4(S_{x}^{2}+S_{z}^{2})}\Big[S\left(S+1\right)-(S_{x}^{2}+S_{z}^{2})\\&-\sqrt{[S\left(S+1\right)-(S_{x}^{2}+S_{z}^{2})]^{2}-(S_{x}^{2}+S_{z}^{2})}\Big].
\end{aligned}
\label{eq:f_SxSz}
\end{align}
This optimization is used to obtain the certified bound for the ratios in Fig.~\ref{fig:ratio_all_m}(b)--(f).
For the optimized measurement angles $\boldsymbol{\theta}^*$ obtained from
the variational ground-state calculation, the lower bound from
Eq.~\eqref{eq:f_SxSz} agrees with the variational value when $N$ becomes large, thereby
certifying the asymptotic quantum value.
\bibliographystyle{apsrev4-2}
\bibliography{references}

\begin{thebibliography}{49}%
\makeatletter
\providecommand \@ifxundefined [1]{%
 \@ifx{#1\undefined}
}%
\providecommand \@ifnum [1]{%
 \ifnum #1\expandafter \@firstoftwo
 \else \expandafter \@secondoftwo
 \fi
}%
\providecommand \@ifx [1]{%
 \ifx #1\expandafter \@firstoftwo
 \else \expandafter \@secondoftwo
 \fi
}%
\providecommand \natexlab [1]{#1}%
\providecommand \enquote  [1]{``#1''}%
\providecommand \bibnamefont  [1]{#1}%
\providecommand \bibfnamefont [1]{#1}%
\providecommand \citenamefont [1]{#1}%
\providecommand \href@noop [0]{\@secondoftwo}%
\providecommand \href [0]{\begingroup \@sanitize@url \@href}%
\providecommand \@href[1]{\@@startlink{#1}\@@href}%
\providecommand \@@href[1]{\endgroup#1\@@endlink}%
\providecommand \@sanitize@url [0]{\catcode `\\12\catcode `\$12\catcode `\&12\catcode `\#12\catcode `\^12\catcode `\_12\catcode `\%12\relax}%
\providecommand \@@startlink[1]{}%
\providecommand \@@endlink[0]{}%
\providecommand \url  [0]{\begingroup\@sanitize@url \@url }%
\providecommand \@url [1]{\endgroup\@href {#1}{\urlprefix }}%
\providecommand \urlprefix  [0]{URL }%
\providecommand \Eprint [0]{\href }%
\providecommand \doibase [0]{https://doi.org/}%
\providecommand \selectlanguage [0]{\@gobble}%
\providecommand \bibinfo  [0]{\@secondoftwo}%
\providecommand \bibfield  [0]{\@secondoftwo}%
\providecommand \translation [1]{[#1]}%
\providecommand \BibitemOpen [0]{}%
\providecommand \bibitemStop [0]{}%
\providecommand \bibitemNoStop [0]{.\EOS\space}%
\providecommand \EOS [0]{\spacefactor3000\relax}%
\providecommand \BibitemShut  [1]{\csname bibitem#1\endcsname}%
\let\auto@bib@innerbib\@empty
%</preamble>
\bibitem [{\citenamefont {Einstein}\ \emph {et~al.}(1935)\citenamefont {Einstein}, \citenamefont {Podolsky},\ and\ \citenamefont {Rosen}}]{einsteinCanQuantumMechanicalDescription1935}%
  \BibitemOpen
  \bibfield  {author} {\bibinfo {author} {\bibfnamefont {A.}~\bibnamefont {Einstein}}, \bibinfo {author} {\bibfnamefont {B.}~\bibnamefont {Podolsky}},\ and\ \bibinfo {author} {\bibfnamefont {N.}~\bibnamefont {Rosen}},\ }\href {https://doi.org/10.1103/PhysRev.47.777} {\bibfield  {journal} {\bibinfo  {journal} {Physical Review}\ }\textbf {\bibinfo {volume} {47}},\ \bibinfo {pages} {777} (\bibinfo {year} {1935})}\BibitemShut {NoStop}%
\bibitem [{\citenamefont {Bohr}(1935)}]{bohrCanQuantumMechanicalDescription1935}%
  \BibitemOpen
  \bibfield  {author} {\bibinfo {author} {\bibfnamefont {N.}~\bibnamefont {Bohr}},\ }\href {https://doi.org/10.1103/PhysRev.48.696} {\bibfield  {journal} {\bibinfo  {journal} {Physical Review}\ }\textbf {\bibinfo {volume} {48}},\ \bibinfo {pages} {696} (\bibinfo {year} {1935})}\BibitemShut {NoStop}%
\bibitem [{\citenamefont {Bell}(1964)}]{bellEinsteinPodolskyRosen1964}%
  \BibitemOpen
  \bibfield  {author} {\bibinfo {author} {\bibfnamefont {J.~S.}\ \bibnamefont {Bell}},\ }\href {https://doi.org/10.1103/PhysicsPhysiqueFizika.1.195} {\bibfield  {journal} {\bibinfo  {journal} {Physics Physique Fizika}\ }\textbf {\bibinfo {volume} {1}},\ \bibinfo {pages} {195} (\bibinfo {year} {1964})}\BibitemShut {NoStop}%
\bibitem [{\citenamefont {Clauser}\ \emph {et~al.}(1969)\citenamefont {Clauser}, \citenamefont {Horne}, \citenamefont {Shimony},\ and\ \citenamefont {Holt}}]{clauserProposedExperimentTest1969}%
  \BibitemOpen
  \bibfield  {author} {\bibinfo {author} {\bibfnamefont {J.~F.}\ \bibnamefont {Clauser}}, \bibinfo {author} {\bibfnamefont {M.~A.}\ \bibnamefont {Horne}}, \bibinfo {author} {\bibfnamefont {A.}~\bibnamefont {Shimony}},\ and\ \bibinfo {author} {\bibfnamefont {R.~A.}\ \bibnamefont {Holt}},\ }\href {https://doi.org/10.1103/PhysRevLett.23.880} {\bibfield  {journal} {\bibinfo  {journal} {Physical Review Letters}\ }\textbf {\bibinfo {volume} {23}},\ \bibinfo {pages} {880} (\bibinfo {year} {1969})}\BibitemShut {NoStop}%
\bibitem [{\citenamefont {Aspect}\ \emph {et~al.}(1982)\citenamefont {Aspect}, \citenamefont {Grangier},\ and\ \citenamefont {Roger}}]{aspectExperimentalRealizationEinsteinPodolskyRosenBohm1982}%
  \BibitemOpen
  \bibfield  {author} {\bibinfo {author} {\bibfnamefont {A.}~\bibnamefont {Aspect}}, \bibinfo {author} {\bibfnamefont {P.}~\bibnamefont {Grangier}},\ and\ \bibinfo {author} {\bibfnamefont {G.}~\bibnamefont {Roger}},\ }\href {https://doi.org/10.1103/PhysRevLett.49.91} {\bibfield  {journal} {\bibinfo  {journal} {Physical Review Letters}\ }\textbf {\bibinfo {volume} {49}},\ \bibinfo {pages} {91} (\bibinfo {year} {1982})}\BibitemShut {NoStop}%
\bibitem [{\citenamefont {Brunner}\ \emph {et~al.}(2014)\citenamefont {Brunner}, \citenamefont {Cavalcanti}, \citenamefont {Pironio}, \citenamefont {Scarani},\ and\ \citenamefont {Wehner}}]{brunnerBellNonlocality2014}%
  \BibitemOpen
  \bibfield  {author} {\bibinfo {author} {\bibfnamefont {N.}~\bibnamefont {Brunner}}, \bibinfo {author} {\bibfnamefont {D.}~\bibnamefont {Cavalcanti}}, \bibinfo {author} {\bibfnamefont {S.}~\bibnamefont {Pironio}}, \bibinfo {author} {\bibfnamefont {V.}~\bibnamefont {Scarani}},\ and\ \bibinfo {author} {\bibfnamefont {S.}~\bibnamefont {Wehner}},\ }\href {https://doi.org/10.1103/RevModPhys.86.419} {\bibfield  {journal} {\bibinfo  {journal} {Reviews of Modern Physics}\ }\textbf {\bibinfo {volume} {86}},\ \bibinfo {pages} {419} (\bibinfo {year} {2014})}\BibitemShut {NoStop}%
\bibitem [{\citenamefont {Fine}(1982)}]{fineHiddenVariablesJoint1982}%
  \BibitemOpen
  \bibfield  {author} {\bibinfo {author} {\bibfnamefont {A.}~\bibnamefont {Fine}},\ }\href {https://doi.org/10.1103/PhysRevLett.48.291} {\bibfield  {journal} {\bibinfo  {journal} {Physical Review Letters}\ }\textbf {\bibinfo {volume} {48}},\ \bibinfo {pages} {291} (\bibinfo {year} {1982})}\BibitemShut {NoStop}%
\bibitem [{\citenamefont {Pitowsky}(1989)}]{pitowskyQuantumProbabilityQuantum1989}%
  \BibitemOpen
  \bibfield  {author} {\bibinfo {author} {\bibfnamefont {I.}~\bibnamefont {Pitowsky}},\ }\href {https://doi.org/10.1007/BFb0021186} {\emph {\bibinfo {title} {Quantum {Probability} — {Quantum} {Logic}}}},\ \bibinfo {series} {Lecture {Notes} in {Physics}}, Vol.\ \bibinfo {volume} {321}\ (\bibinfo  {publisher} {Springer-Verlag},\ \bibinfo {address} {Berlin/Heidelberg},\ \bibinfo {year} {1989})\BibitemShut {NoStop}%
\bibitem [{\citenamefont {Hensen}\ \emph {et~al.}(2015)\citenamefont {Hensen}, \citenamefont {Bernien}, \citenamefont {Dréau}, \citenamefont {Reiserer}, \citenamefont {Kalb}, \citenamefont {Blok}, \citenamefont {Ruitenberg}, \citenamefont {Vermeulen}, \citenamefont {Schouten}, \citenamefont {Abellán}, \citenamefont {Amaya}, \citenamefont {Pruneri}, \citenamefont {Mitchell}, \citenamefont {Markham}, \citenamefont {Twitchen}, \citenamefont {Elkouss}, \citenamefont {Wehner}, \citenamefont {Taminiau},\ and\ \citenamefont {Hanson}}]{hensenLoopholefreeBellInequality2015}%
  \BibitemOpen
  \bibfield  {author} {\bibinfo {author} {\bibfnamefont {B.}~\bibnamefont {Hensen}}, \bibinfo {author} {\bibfnamefont {H.}~\bibnamefont {Bernien}}, \bibinfo {author} {\bibfnamefont {A.~E.}\ \bibnamefont {Dréau}}, \bibinfo {author} {\bibfnamefont {A.}~\bibnamefont {Reiserer}}, \bibinfo {author} {\bibfnamefont {N.}~\bibnamefont {Kalb}}, \bibinfo {author} {\bibfnamefont {M.~S.}\ \bibnamefont {Blok}}, \bibinfo {author} {\bibfnamefont {J.}~\bibnamefont {Ruitenberg}}, \bibinfo {author} {\bibfnamefont {R.~F.~L.}\ \bibnamefont {Vermeulen}}, \bibinfo {author} {\bibfnamefont {R.~N.}\ \bibnamefont {Schouten}}, \bibinfo {author} {\bibfnamefont {C.}~\bibnamefont {Abellán}}, \bibinfo {author} {\bibfnamefont {W.}~\bibnamefont {Amaya}}, \bibinfo {author} {\bibfnamefont {V.}~\bibnamefont {Pruneri}}, \bibinfo {author} {\bibfnamefont {M.~W.}\ \bibnamefont {Mitchell}}, \bibinfo {author} {\bibfnamefont {M.}~\bibnamefont {Markham}}, \bibinfo {author} {\bibfnamefont {D.~J.}\ \bibnamefont {Twitchen}}, \bibinfo {author}
  {\bibfnamefont {D.}~\bibnamefont {Elkouss}}, \bibinfo {author} {\bibfnamefont {S.}~\bibnamefont {Wehner}}, \bibinfo {author} {\bibfnamefont {T.~H.}\ \bibnamefont {Taminiau}},\ and\ \bibinfo {author} {\bibfnamefont {R.}~\bibnamefont {Hanson}},\ }\href {https://doi.org/10.1038/nature15759} {\bibfield  {journal} {\bibinfo  {journal} {Nature}\ }\textbf {\bibinfo {volume} {526}},\ \bibinfo {pages} {682} (\bibinfo {year} {2015})}\BibitemShut {NoStop}%
\bibitem [{\citenamefont {Giustina}\ \emph {et~al.}(2015)\citenamefont {Giustina}, \citenamefont {Versteegh}, \citenamefont {Wengerowsky}, \citenamefont {Handsteiner}, \citenamefont {Hochrainer}, \citenamefont {Phelan}, \citenamefont {Steinlechner}, \citenamefont {Kofler}, \citenamefont {Larsson}, \citenamefont {Abellán}, \citenamefont {Amaya}, \citenamefont {Pruneri}, \citenamefont {Mitchell}, \citenamefont {Beyer}, \citenamefont {Gerrits}, \citenamefont {Lita}, \citenamefont {Shalm}, \citenamefont {Nam}, \citenamefont {Scheidl}, \citenamefont {Ursin}, \citenamefont {Wittmann},\ and\ \citenamefont {Zeilinger}}]{giustinaSignificantLoopholeFreeTestBells2015}%
  \BibitemOpen
  \bibfield  {author} {\bibinfo {author} {\bibfnamefont {M.}~\bibnamefont {Giustina}}, \bibinfo {author} {\bibfnamefont {M.~A.~M.}\ \bibnamefont {Versteegh}}, \bibinfo {author} {\bibfnamefont {S.}~\bibnamefont {Wengerowsky}}, \bibinfo {author} {\bibfnamefont {J.}~\bibnamefont {Handsteiner}}, \bibinfo {author} {\bibfnamefont {A.}~\bibnamefont {Hochrainer}}, \bibinfo {author} {\bibfnamefont {K.}~\bibnamefont {Phelan}}, \bibinfo {author} {\bibfnamefont {F.}~\bibnamefont {Steinlechner}}, \bibinfo {author} {\bibfnamefont {J.}~\bibnamefont {Kofler}}, \bibinfo {author} {\bibfnamefont {J.-A.}\ \bibnamefont {Larsson}}, \bibinfo {author} {\bibfnamefont {C.}~\bibnamefont {Abellán}}, \bibinfo {author} {\bibfnamefont {W.}~\bibnamefont {Amaya}}, \bibinfo {author} {\bibfnamefont {V.}~\bibnamefont {Pruneri}}, \bibinfo {author} {\bibfnamefont {M.~W.}\ \bibnamefont {Mitchell}}, \bibinfo {author} {\bibfnamefont {J.}~\bibnamefont {Beyer}}, \bibinfo {author} {\bibfnamefont {T.}~\bibnamefont {Gerrits}}, \bibinfo {author}
  {\bibfnamefont {A.~E.}\ \bibnamefont {Lita}}, \bibinfo {author} {\bibfnamefont {L.~K.}\ \bibnamefont {Shalm}}, \bibinfo {author} {\bibfnamefont {S.~W.}\ \bibnamefont {Nam}}, \bibinfo {author} {\bibfnamefont {T.}~\bibnamefont {Scheidl}}, \bibinfo {author} {\bibfnamefont {R.}~\bibnamefont {Ursin}}, \bibinfo {author} {\bibfnamefont {B.}~\bibnamefont {Wittmann}},\ and\ \bibinfo {author} {\bibfnamefont {A.}~\bibnamefont {Zeilinger}},\ }\href {https://doi.org/10.1103/PhysRevLett.115.250401} {\bibfield  {journal} {\bibinfo  {journal} {Physical Review Letters}\ }\textbf {\bibinfo {volume} {115}},\ \bibinfo {pages} {250401} (\bibinfo {year} {2015})}\BibitemShut {NoStop}%
\bibitem [{\citenamefont {Shalm}\ \emph {et~al.}(2015)\citenamefont {Shalm}, \citenamefont {Meyer-Scott}, \citenamefont {Christensen}, \citenamefont {Bierhorst}, \citenamefont {Wayne}, \citenamefont {Stevens}, \citenamefont {Gerrits}, \citenamefont {Glancy}, \citenamefont {Hamel}, \citenamefont {Allman}, \citenamefont {Coakley}, \citenamefont {Dyer}, \citenamefont {Hodge}, \citenamefont {Lita}, \citenamefont {Verma}, \citenamefont {Lambrocco}, \citenamefont {Tortorici}, \citenamefont {Migdall}, \citenamefont {Zhang}, \citenamefont {Kumor}, \citenamefont {Farr}, \citenamefont {Marsili}, \citenamefont {Shaw}, \citenamefont {Stern}, \citenamefont {Abellán}, \citenamefont {Amaya}, \citenamefont {Pruneri}, \citenamefont {Jennewein}, \citenamefont {Mitchell}, \citenamefont {Kwiat}, \citenamefont {Bienfang}, \citenamefont {Mirin}, \citenamefont {Knill},\ and\ \citenamefont {Nam}}]{shalmStrongLoopholeFreeTest2015}%
  \BibitemOpen
  \bibfield  {author} {\bibinfo {author} {\bibfnamefont {L.~K.}\ \bibnamefont {Shalm}}, \bibinfo {author} {\bibfnamefont {E.}~\bibnamefont {Meyer-Scott}}, \bibinfo {author} {\bibfnamefont {B.~G.}\ \bibnamefont {Christensen}}, \bibinfo {author} {\bibfnamefont {P.}~\bibnamefont {Bierhorst}}, \bibinfo {author} {\bibfnamefont {M.~A.}\ \bibnamefont {Wayne}}, \bibinfo {author} {\bibfnamefont {M.~J.}\ \bibnamefont {Stevens}}, \bibinfo {author} {\bibfnamefont {T.}~\bibnamefont {Gerrits}}, \bibinfo {author} {\bibfnamefont {S.}~\bibnamefont {Glancy}}, \bibinfo {author} {\bibfnamefont {D.~R.}\ \bibnamefont {Hamel}}, \bibinfo {author} {\bibfnamefont {M.~S.}\ \bibnamefont {Allman}}, \bibinfo {author} {\bibfnamefont {K.~J.}\ \bibnamefont {Coakley}}, \bibinfo {author} {\bibfnamefont {S.~D.}\ \bibnamefont {Dyer}}, \bibinfo {author} {\bibfnamefont {C.}~\bibnamefont {Hodge}}, \bibinfo {author} {\bibfnamefont {A.~E.}\ \bibnamefont {Lita}}, \bibinfo {author} {\bibfnamefont {V.~B.}\ \bibnamefont {Verma}}, \bibinfo {author}
  {\bibfnamefont {C.}~\bibnamefont {Lambrocco}}, \bibinfo {author} {\bibfnamefont {E.}~\bibnamefont {Tortorici}}, \bibinfo {author} {\bibfnamefont {A.~L.}\ \bibnamefont {Migdall}}, \bibinfo {author} {\bibfnamefont {Y.}~\bibnamefont {Zhang}}, \bibinfo {author} {\bibfnamefont {D.~R.}\ \bibnamefont {Kumor}}, \bibinfo {author} {\bibfnamefont {W.~H.}\ \bibnamefont {Farr}}, \bibinfo {author} {\bibfnamefont {F.}~\bibnamefont {Marsili}}, \bibinfo {author} {\bibfnamefont {M.~D.}\ \bibnamefont {Shaw}}, \bibinfo {author} {\bibfnamefont {J.~A.}\ \bibnamefont {Stern}}, \bibinfo {author} {\bibfnamefont {C.}~\bibnamefont {Abellán}}, \bibinfo {author} {\bibfnamefont {W.}~\bibnamefont {Amaya}}, \bibinfo {author} {\bibfnamefont {V.}~\bibnamefont {Pruneri}}, \bibinfo {author} {\bibfnamefont {T.}~\bibnamefont {Jennewein}}, \bibinfo {author} {\bibfnamefont {M.~W.}\ \bibnamefont {Mitchell}}, \bibinfo {author} {\bibfnamefont {P.~G.}\ \bibnamefont {Kwiat}}, \bibinfo {author} {\bibfnamefont {J.~C.}\ \bibnamefont {Bienfang}},
  \bibinfo {author} {\bibfnamefont {R.~P.}\ \bibnamefont {Mirin}}, \bibinfo {author} {\bibfnamefont {E.}~\bibnamefont {Knill}},\ and\ \bibinfo {author} {\bibfnamefont {S.~W.}\ \bibnamefont {Nam}},\ }\href {https://doi.org/10.1103/PhysRevLett.115.250402} {\bibfield  {journal} {\bibinfo  {journal} {Physical Review Letters}\ }\textbf {\bibinfo {volume} {115}},\ \bibinfo {pages} {250402} (\bibinfo {year} {2015})}\BibitemShut {NoStop}%
\bibitem [{\citenamefont {Rosenfeld}\ \emph {et~al.}(2017)\citenamefont {Rosenfeld}, \citenamefont {Burchardt}, \citenamefont {Garthoff}, \citenamefont {Redeker}, \citenamefont {Ortegel}, \citenamefont {Rau},\ and\ \citenamefont {Weinfurter}}]{rosenfeldEventReadyBellTest2017}%
  \BibitemOpen
  \bibfield  {author} {\bibinfo {author} {\bibfnamefont {W.}~\bibnamefont {Rosenfeld}}, \bibinfo {author} {\bibfnamefont {D.}~\bibnamefont {Burchardt}}, \bibinfo {author} {\bibfnamefont {R.}~\bibnamefont {Garthoff}}, \bibinfo {author} {\bibfnamefont {K.}~\bibnamefont {Redeker}}, \bibinfo {author} {\bibfnamefont {N.}~\bibnamefont {Ortegel}}, \bibinfo {author} {\bibfnamefont {M.}~\bibnamefont {Rau}},\ and\ \bibinfo {author} {\bibfnamefont {H.}~\bibnamefont {Weinfurter}},\ }\href {https://doi.org/10.1103/PhysRevLett.119.010402} {\bibfield  {journal} {\bibinfo  {journal} {Physical Review Letters}\ }\textbf {\bibinfo {volume} {119}},\ \bibinfo {pages} {010402} (\bibinfo {year} {2017})}\BibitemShut {NoStop}%
\bibitem [{\citenamefont {Acín}\ \emph {et~al.}(2007)\citenamefont {Acín}, \citenamefont {Brunner}, \citenamefont {Gisin}, \citenamefont {Massar}, \citenamefont {Pironio},\ and\ \citenamefont {Scarani}}]{acinDeviceIndependentSecurityQuantum2007}%
  \BibitemOpen
  \bibfield  {author} {\bibinfo {author} {\bibfnamefont {A.}~\bibnamefont {Acín}}, \bibinfo {author} {\bibfnamefont {N.}~\bibnamefont {Brunner}}, \bibinfo {author} {\bibfnamefont {N.}~\bibnamefont {Gisin}}, \bibinfo {author} {\bibfnamefont {S.}~\bibnamefont {Massar}}, \bibinfo {author} {\bibfnamefont {S.}~\bibnamefont {Pironio}},\ and\ \bibinfo {author} {\bibfnamefont {V.}~\bibnamefont {Scarani}},\ }\href {https://doi.org/10.1103/PhysRevLett.98.230501} {\bibfield  {journal} {\bibinfo  {journal} {Physical Review Letters}\ }\textbf {\bibinfo {volume} {98}},\ \bibinfo {pages} {230501} (\bibinfo {year} {2007})}\BibitemShut {NoStop}%
\bibitem [{\citenamefont {Pironio}\ \emph {et~al.}(2009)\citenamefont {Pironio}, \citenamefont {Acín}, \citenamefont {Brunner}, \citenamefont {Gisin}, \citenamefont {Massar},\ and\ \citenamefont {Scarani}}]{pironioDeviceindependentQuantumKey2009}%
  \BibitemOpen
  \bibfield  {author} {\bibinfo {author} {\bibfnamefont {S.}~\bibnamefont {Pironio}}, \bibinfo {author} {\bibfnamefont {A.}~\bibnamefont {Acín}}, \bibinfo {author} {\bibfnamefont {N.}~\bibnamefont {Brunner}}, \bibinfo {author} {\bibfnamefont {N.}~\bibnamefont {Gisin}}, \bibinfo {author} {\bibfnamefont {S.}~\bibnamefont {Massar}},\ and\ \bibinfo {author} {\bibfnamefont {V.}~\bibnamefont {Scarani}},\ }\href {https://doi.org/10.1088/1367-2630/11/4/045021} {\bibfield  {journal} {\bibinfo  {journal} {New Journal of Physics}\ }\textbf {\bibinfo {volume} {11}},\ \bibinfo {pages} {045021} (\bibinfo {year} {2009})}\BibitemShut {NoStop}%
\bibitem [{\citenamefont {Vazirani}\ and\ \citenamefont {Vidick}(2014)}]{vaziraniFullyDeviceIndependentQuantum2014}%
  \BibitemOpen
  \bibfield  {author} {\bibinfo {author} {\bibfnamefont {U.}~\bibnamefont {Vazirani}}\ and\ \bibinfo {author} {\bibfnamefont {T.}~\bibnamefont {Vidick}},\ }\href {https://doi.org/10.1103/PhysRevLett.113.140501} {\bibfield  {journal} {\bibinfo  {journal} {Physical Review Letters}\ }\textbf {\bibinfo {volume} {113}},\ \bibinfo {pages} {140501} (\bibinfo {year} {2014})}\BibitemShut {NoStop}%
\bibitem [{\citenamefont {Xu}\ \emph {et~al.}(2020)\citenamefont {Xu}, \citenamefont {Ma}, \citenamefont {Zhang}, \citenamefont {Lo},\ and\ \citenamefont {Pan}}]{xuSecureQuantumKey2020}%
  \BibitemOpen
  \bibfield  {author} {\bibinfo {author} {\bibfnamefont {F.}~\bibnamefont {Xu}}, \bibinfo {author} {\bibfnamefont {X.}~\bibnamefont {Ma}}, \bibinfo {author} {\bibfnamefont {Q.}~\bibnamefont {Zhang}}, \bibinfo {author} {\bibfnamefont {H.-K.}\ \bibnamefont {Lo}},\ and\ \bibinfo {author} {\bibfnamefont {J.-W.}\ \bibnamefont {Pan}},\ }\href {https://doi.org/10.1103/RevModPhys.92.025002} {\bibfield  {journal} {\bibinfo  {journal} {Reviews of Modern Physics}\ }\textbf {\bibinfo {volume} {92}},\ \bibinfo {pages} {025002} (\bibinfo {year} {2020})}\BibitemShut {NoStop}%
\bibitem [{\citenamefont {Pironio}\ \emph {et~al.}(2010)\citenamefont {Pironio}, \citenamefont {Acín}, \citenamefont {Massar}, \citenamefont {De~La~Giroday}, \citenamefont {Matsukevich}, \citenamefont {Maunz}, \citenamefont {Olmschenk}, \citenamefont {Hayes}, \citenamefont {Luo}, \citenamefont {Manning},\ and\ \citenamefont {Monroe}}]{pironioRandomNumbersCertified2010}%
  \BibitemOpen
  \bibfield  {author} {\bibinfo {author} {\bibfnamefont {S.}~\bibnamefont {Pironio}}, \bibinfo {author} {\bibfnamefont {A.}~\bibnamefont {Acín}}, \bibinfo {author} {\bibfnamefont {S.}~\bibnamefont {Massar}}, \bibinfo {author} {\bibfnamefont {A.~B.}\ \bibnamefont {De~La~Giroday}}, \bibinfo {author} {\bibfnamefont {D.~N.}\ \bibnamefont {Matsukevich}}, \bibinfo {author} {\bibfnamefont {P.}~\bibnamefont {Maunz}}, \bibinfo {author} {\bibfnamefont {S.}~\bibnamefont {Olmschenk}}, \bibinfo {author} {\bibfnamefont {D.}~\bibnamefont {Hayes}}, \bibinfo {author} {\bibfnamefont {L.}~\bibnamefont {Luo}}, \bibinfo {author} {\bibfnamefont {T.~A.}\ \bibnamefont {Manning}},\ and\ \bibinfo {author} {\bibfnamefont {C.}~\bibnamefont {Monroe}},\ }\href {https://doi.org/10.1038/nature09008} {\bibfield  {journal} {\bibinfo  {journal} {Nature}\ }\textbf {\bibinfo {volume} {464}},\ \bibinfo {pages} {1021} (\bibinfo {year} {2010})}\BibitemShut {NoStop}%
\bibitem [{\citenamefont {Fyrillas}\ \emph {et~al.}(2024)\citenamefont {Fyrillas}, \citenamefont {Bourdoncle}, \citenamefont {Maïnos}, \citenamefont {Emeriau}, \citenamefont {Start}, \citenamefont {Margaria}, \citenamefont {Morassi}, \citenamefont {Lemaître}, \citenamefont {Sagnes}, \citenamefont {Stepanov}, \citenamefont {Au}, \citenamefont {Boissier}, \citenamefont {Somaschi}, \citenamefont {Maring}, \citenamefont {Belabas},\ and\ \citenamefont {Mansfield}}]{fyrillasCertifiedRandomnessTight2024}%
  \BibitemOpen
  \bibfield  {author} {\bibinfo {author} {\bibfnamefont {A.}~\bibnamefont {Fyrillas}}, \bibinfo {author} {\bibfnamefont {B.}~\bibnamefont {Bourdoncle}}, \bibinfo {author} {\bibfnamefont {A.}~\bibnamefont {Maïnos}}, \bibinfo {author} {\bibfnamefont {P.-E.}\ \bibnamefont {Emeriau}}, \bibinfo {author} {\bibfnamefont {K.}~\bibnamefont {Start}}, \bibinfo {author} {\bibfnamefont {N.}~\bibnamefont {Margaria}}, \bibinfo {author} {\bibfnamefont {M.}~\bibnamefont {Morassi}}, \bibinfo {author} {\bibfnamefont {A.}~\bibnamefont {Lemaître}}, \bibinfo {author} {\bibfnamefont {I.}~\bibnamefont {Sagnes}}, \bibinfo {author} {\bibfnamefont {P.}~\bibnamefont {Stepanov}}, \bibinfo {author} {\bibfnamefont {T.~H.}\ \bibnamefont {Au}}, \bibinfo {author} {\bibfnamefont {S.}~\bibnamefont {Boissier}}, \bibinfo {author} {\bibfnamefont {N.}~\bibnamefont {Somaschi}}, \bibinfo {author} {\bibfnamefont {N.}~\bibnamefont {Maring}}, \bibinfo {author} {\bibfnamefont {N.}~\bibnamefont {Belabas}},\ and\ \bibinfo {author} {\bibfnamefont
  {S.}~\bibnamefont {Mansfield}},\ }\href {https://doi.org/10.1103/PRXQuantum.5.020348} {\bibfield  {journal} {\bibinfo  {journal} {PRX Quantum}\ }\textbf {\bibinfo {volume} {5}},\ \bibinfo {pages} {020348} (\bibinfo {year} {2024})}\BibitemShut {NoStop}%
\bibitem [{\citenamefont {Zhang}\ \emph {et~al.}(2025)\citenamefont {Zhang}, \citenamefont {Li}, \citenamefont {Hu}, \citenamefont {Xiang}, \citenamefont {Li}, \citenamefont {Guo}, \citenamefont {Tura}, \citenamefont {Gong}, \citenamefont {He},\ and\ \citenamefont {Liu}}]{zhangRandomnessNonlocalityMultipleInput2025}%
  \BibitemOpen
  \bibfield  {author} {\bibinfo {author} {\bibfnamefont {C.}~\bibnamefont {Zhang}}, \bibinfo {author} {\bibfnamefont {Y.}~\bibnamefont {Li}}, \bibinfo {author} {\bibfnamefont {X.-M.}\ \bibnamefont {Hu}}, \bibinfo {author} {\bibfnamefont {Y.}~\bibnamefont {Xiang}}, \bibinfo {author} {\bibfnamefont {C.-F.}\ \bibnamefont {Li}}, \bibinfo {author} {\bibfnamefont {G.-C.}\ \bibnamefont {Guo}}, \bibinfo {author} {\bibfnamefont {J.}~\bibnamefont {Tura}}, \bibinfo {author} {\bibfnamefont {Q.}~\bibnamefont {Gong}}, \bibinfo {author} {\bibfnamefont {Q.}~\bibnamefont {He}},\ and\ \bibinfo {author} {\bibfnamefont {B.-H.}\ \bibnamefont {Liu}},\ }\href {https://doi.org/10.1103/PhysRevLett.134.090201} {\bibfield  {journal} {\bibinfo  {journal} {Physical Review Letters}\ }\textbf {\bibinfo {volume} {134}},\ \bibinfo {pages} {090201} (\bibinfo {year} {2025})}\BibitemShut {NoStop}%
\bibitem [{\citenamefont {Šupić}\ and\ \citenamefont {Bowles}(2020)}]{supicSelftestingQuantumSystems2020b}%
  \BibitemOpen
  \bibfield  {author} {\bibinfo {author} {\bibfnamefont {I.}~\bibnamefont {Šupić}}\ and\ \bibinfo {author} {\bibfnamefont {J.}~\bibnamefont {Bowles}},\ }\href {https://doi.org/10.22331/q-2020-09-30-337} {\bibfield  {journal} {\bibinfo  {journal} {Quantum}\ }\textbf {\bibinfo {volume} {4}},\ \bibinfo {pages} {337} (\bibinfo {year} {2020})}\BibitemShut {NoStop}%
\bibitem [{\citenamefont {Li}\ \emph {et~al.}(2025)\citenamefont {Li}, \citenamefont {Xiang}, \citenamefont {Tura},\ and\ \citenamefont {He}}]{liNecessarySufficientCondition2025}%
  \BibitemOpen
  \bibfield  {author} {\bibinfo {author} {\bibfnamefont {Y.}~\bibnamefont {Li}}, \bibinfo {author} {\bibfnamefont {Y.}~\bibnamefont {Xiang}}, \bibinfo {author} {\bibfnamefont {J.}~\bibnamefont {Tura}},\ and\ \bibinfo {author} {\bibfnamefont {Q.}~\bibnamefont {He}},\ }\href {https://doi.org/10.1103/zp36-2xg4} {\bibfield  {journal} {\bibinfo  {journal} {Physical Review Letters}\ }\textbf {\bibinfo {volume} {135}},\ \bibinfo {pages} {060201} (\bibinfo {year} {2025})}\BibitemShut {NoStop}%
\bibitem [{\citenamefont {Baccari}\ \emph {et~al.}(2020)\citenamefont {Baccari}, \citenamefont {Augusiak}, \citenamefont {Šupić}, \citenamefont {Tura},\ and\ \citenamefont {Acín}}]{baccariScalableBellInequalities2020}%
  \BibitemOpen
  \bibfield  {author} {\bibinfo {author} {\bibfnamefont {F.}~\bibnamefont {Baccari}}, \bibinfo {author} {\bibfnamefont {R.}~\bibnamefont {Augusiak}}, \bibinfo {author} {\bibfnamefont {I.}~\bibnamefont {Šupić}}, \bibinfo {author} {\bibfnamefont {J.}~\bibnamefont {Tura}},\ and\ \bibinfo {author} {\bibfnamefont {A.}~\bibnamefont {Acín}},\ }\href {https://doi.org/10.1103/PhysRevLett.124.020402} {\bibfield  {journal} {\bibinfo  {journal} {Physical Review Letters}\ }\textbf {\bibinfo {volume} {124}},\ \bibinfo {pages} {020402} (\bibinfo {year} {2020})}\BibitemShut {NoStop}%
\bibitem [{\citenamefont {Mermin}(1990)}]{merminExtremeQuantumEntanglement1990}%
  \BibitemOpen
  \bibfield  {author} {\bibinfo {author} {\bibfnamefont {N.~D.}\ \bibnamefont {Mermin}},\ }\href {https://doi.org/10.1103/PhysRevLett.65.1838} {\bibfield  {journal} {\bibinfo  {journal} {Physical Review Letters}\ }\textbf {\bibinfo {volume} {65}},\ \bibinfo {pages} {1838} (\bibinfo {year} {1990})}\BibitemShut {NoStop}%
\bibitem [{\citenamefont {Svetlichny}(1987)}]{svetlichnyDistinguishingThreebodyTwobody1987}%
  \BibitemOpen
  \bibfield  {author} {\bibinfo {author} {\bibfnamefont {G.}~\bibnamefont {Svetlichny}},\ }\href {https://doi.org/10.1103/PhysRevD.35.3066} {\bibfield  {journal} {\bibinfo  {journal} {Physical Review D}\ }\textbf {\bibinfo {volume} {35}},\ \bibinfo {pages} {3066} (\bibinfo {year} {1987})}\BibitemShut {NoStop}%
\bibitem [{\citenamefont {Bancal}\ \emph {et~al.}(2011{\natexlab{a}})\citenamefont {Bancal}, \citenamefont {Gisin}, \citenamefont {Liang},\ and\ \citenamefont {Pironio}}]{bancalDeviceIndependentWitnessesGenuine2011}%
  \BibitemOpen
  \bibfield  {author} {\bibinfo {author} {\bibfnamefont {J.-D.}\ \bibnamefont {Bancal}}, \bibinfo {author} {\bibfnamefont {N.}~\bibnamefont {Gisin}}, \bibinfo {author} {\bibfnamefont {Y.-C.}\ \bibnamefont {Liang}},\ and\ \bibinfo {author} {\bibfnamefont {S.}~\bibnamefont {Pironio}},\ }\href {https://doi.org/10.1103/PhysRevLett.106.250404} {\bibfield  {journal} {\bibinfo  {journal} {Physical Review Letters}\ }\textbf {\bibinfo {volume} {106}},\ \bibinfo {pages} {250404} (\bibinfo {year} {2011}{\natexlab{a}})}\BibitemShut {NoStop}%
\bibitem [{\citenamefont {Pitowsky}(1991)}]{pitowskyCorrelationPolytopesTheir1991}%
  \BibitemOpen
  \bibfield  {author} {\bibinfo {author} {\bibfnamefont {I.}~\bibnamefont {Pitowsky}},\ }\href@noop {} {\bibfield  {journal} {\bibinfo  {journal} {Mathematical Programming}\ }\textbf {\bibinfo {volume} {50}},\ \bibinfo {pages} {395} (\bibinfo {year} {1991})}\BibitemShut {NoStop}%
\bibitem [{\citenamefont {Zukowski}\ and\ \citenamefont {Brukner}(2002)}]{zukowskiBellsTheoremGeneral2002}%
  \BibitemOpen
  \bibfield  {author} {\bibinfo {author} {\bibfnamefont {M.}~\bibnamefont {Zukowski}}\ and\ \bibinfo {author} {\bibfnamefont {C.}~\bibnamefont {Brukner}},\ }\href {https://doi.org/10.1103/PhysRevLett.88.210401} {\bibfield  {journal} {\bibinfo  {journal} {Physical Review Letters}\ }\textbf {\bibinfo {volume} {88}},\ \bibinfo {pages} {210401} (\bibinfo {year} {2002})}\BibitemShut {NoStop}%
\bibitem [{\citenamefont {Bancal}\ \emph {et~al.}(2011{\natexlab{b}})\citenamefont {Bancal}, \citenamefont {Brunner}, \citenamefont {Gisin},\ and\ \citenamefont {Liang}}]{bancalDetectingGenuineMultipartite2011}%
  \BibitemOpen
  \bibfield  {author} {\bibinfo {author} {\bibfnamefont {J.-D.}\ \bibnamefont {Bancal}}, \bibinfo {author} {\bibfnamefont {N.}~\bibnamefont {Brunner}}, \bibinfo {author} {\bibfnamefont {N.}~\bibnamefont {Gisin}},\ and\ \bibinfo {author} {\bibfnamefont {Y.-C.}\ \bibnamefont {Liang}},\ }\href {https://doi.org/10.1103/PhysRevLett.106.020405} {\bibfield  {journal} {\bibinfo  {journal} {Physical Review Letters}\ }\textbf {\bibinfo {volume} {106}},\ \bibinfo {pages} {020405} (\bibinfo {year} {2011}{\natexlab{b}})}\BibitemShut {NoStop}%
\bibitem [{\citenamefont {Werner}\ and\ \citenamefont {Wolf}(2001)}]{wernerAllmultipartiteBellcorrelationInequalities2001}%
  \BibitemOpen
  \bibfield  {author} {\bibinfo {author} {\bibfnamefont {R.~F.}\ \bibnamefont {Werner}}\ and\ \bibinfo {author} {\bibfnamefont {M.~M.}\ \bibnamefont {Wolf}},\ }\href {https://doi.org/10.1103/PhysRevA.64.032112} {\bibfield  {journal} {\bibinfo  {journal} {Physical Review A}\ }\textbf {\bibinfo {volume} {64}},\ \bibinfo {pages} {032112} (\bibinfo {year} {2001})}\BibitemShut {NoStop}%
\bibitem [{\citenamefont {Collins}\ \emph {et~al.}(2002)\citenamefont {Collins}, \citenamefont {Gisin}, \citenamefont {Linden}, \citenamefont {Massar},\ and\ \citenamefont {Popescu}}]{collinsBellInequalitiesArbitrarily2002a}%
  \BibitemOpen
  \bibfield  {author} {\bibinfo {author} {\bibfnamefont {D.}~\bibnamefont {Collins}}, \bibinfo {author} {\bibfnamefont {N.}~\bibnamefont {Gisin}}, \bibinfo {author} {\bibfnamefont {N.}~\bibnamefont {Linden}}, \bibinfo {author} {\bibfnamefont {S.}~\bibnamefont {Massar}},\ and\ \bibinfo {author} {\bibfnamefont {S.}~\bibnamefont {Popescu}},\ }\href {https://doi.org/10.1103/PhysRevLett.88.040404} {\bibfield  {journal} {\bibinfo  {journal} {Physical Review Letters}\ }\textbf {\bibinfo {volume} {88}},\ \bibinfo {pages} {040404} (\bibinfo {year} {2002})}\BibitemShut {NoStop}%
\bibitem [{\citenamefont {Tura}\ \emph {et~al.}(2014)\citenamefont {Tura}, \citenamefont {Augusiak}, \citenamefont {Sainz}, \citenamefont {Vértesi}, \citenamefont {Lewenstein},\ and\ \citenamefont {Acín}}]{turaDetectingNonlocalityManybody2014c}%
  \BibitemOpen
  \bibfield  {author} {\bibinfo {author} {\bibfnamefont {J.}~\bibnamefont {Tura}}, \bibinfo {author} {\bibfnamefont {R.}~\bibnamefont {Augusiak}}, \bibinfo {author} {\bibfnamefont {A.~B.}\ \bibnamefont {Sainz}}, \bibinfo {author} {\bibfnamefont {T.}~\bibnamefont {Vértesi}}, \bibinfo {author} {\bibfnamefont {M.}~\bibnamefont {Lewenstein}},\ and\ \bibinfo {author} {\bibfnamefont {A.}~\bibnamefont {Acín}},\ }\href {https://doi.org/10.1126/science.1247715} {\bibfield  {journal} {\bibinfo  {journal} {Science}\ }\textbf {\bibinfo {volume} {344}},\ \bibinfo {pages} {1256} (\bibinfo {year} {2014})}\BibitemShut {NoStop}%
\bibitem [{\citenamefont {Tura}\ \emph {et~al.}(2015)\citenamefont {Tura}, \citenamefont {Augusiak}, \citenamefont {Sainz}, \citenamefont {Lücke}, \citenamefont {Klempt}, \citenamefont {Lewenstein},\ and\ \citenamefont {Acín}}]{turaNonlocalityManybodyQuantum2015}%
  \BibitemOpen
  \bibfield  {author} {\bibinfo {author} {\bibfnamefont {J.}~\bibnamefont {Tura}}, \bibinfo {author} {\bibfnamefont {R.}~\bibnamefont {Augusiak}}, \bibinfo {author} {\bibfnamefont {A.~B.}\ \bibnamefont {Sainz}}, \bibinfo {author} {\bibfnamefont {B.}~\bibnamefont {Lücke}}, \bibinfo {author} {\bibfnamefont {C.}~\bibnamefont {Klempt}}, \bibinfo {author} {\bibfnamefont {M.}~\bibnamefont {Lewenstein}},\ and\ \bibinfo {author} {\bibfnamefont {A.}~\bibnamefont {Acín}},\ }\href {https://doi.org/10.1016/j.aop.2015.07.021} {\bibfield  {journal} {\bibinfo  {journal} {Annals of Physics}\ }\textbf {\bibinfo {volume} {362}},\ \bibinfo {pages} {370} (\bibinfo {year} {2015})}\BibitemShut {NoStop}%
\bibitem [{\citenamefont {Wagner}\ \emph {et~al.}(2017)\citenamefont {Wagner}, \citenamefont {Schmied}, \citenamefont {Fadel}, \citenamefont {Treutlein}, \citenamefont {Sangouard},\ and\ \citenamefont {Bancal}}]{wagnerBellCorrelationsManyBody2017}%
  \BibitemOpen
  \bibfield  {author} {\bibinfo {author} {\bibfnamefont {S.}~\bibnamefont {Wagner}}, \bibinfo {author} {\bibfnamefont {R.}~\bibnamefont {Schmied}}, \bibinfo {author} {\bibfnamefont {M.}~\bibnamefont {Fadel}}, \bibinfo {author} {\bibfnamefont {P.}~\bibnamefont {Treutlein}}, \bibinfo {author} {\bibfnamefont {N.}~\bibnamefont {Sangouard}},\ and\ \bibinfo {author} {\bibfnamefont {J.-D.}\ \bibnamefont {Bancal}},\ }\href {https://doi.org/10.1103/PhysRevLett.119.170403} {\bibfield  {journal} {\bibinfo  {journal} {Physical Review Letters}\ }\textbf {\bibinfo {volume} {119}},\ \bibinfo {pages} {170403} (\bibinfo {year} {2017})}\BibitemShut {NoStop}%
\bibitem [{\citenamefont {Müller-Rigat}\ \emph {et~al.}(2021)\citenamefont {Müller-Rigat}, \citenamefont {Aloy}, \citenamefont {Lewenstein},\ and\ \citenamefont {Frérot}}]{muller-rigatInferringNonlinearManyBody2021}%
  \BibitemOpen
  \bibfield  {author} {\bibinfo {author} {\bibfnamefont {G.}~\bibnamefont {Müller-Rigat}}, \bibinfo {author} {\bibfnamefont {A.}~\bibnamefont {Aloy}}, \bibinfo {author} {\bibfnamefont {M.}~\bibnamefont {Lewenstein}},\ and\ \bibinfo {author} {\bibfnamefont {I.}~\bibnamefont {Frérot}},\ }\href {https://doi.org/10.1103/PRXQuantum.2.030329} {\bibfield  {journal} {\bibinfo  {journal} {PRX Quantum}\ }\textbf {\bibinfo {volume} {2}},\ \bibinfo {pages} {030329} (\bibinfo {year} {2021})},\ \bibinfo {note} {arXiv:2012.08474 [quant-ph]}\BibitemShut {NoStop}%
\bibitem [{\citenamefont {Guo}\ \emph {et~al.}(2023)\citenamefont {Guo}, \citenamefont {Tura}, \citenamefont {He},\ and\ \citenamefont {Fadel}}]{guoDetectingBellCorrelations2023a}%
  \BibitemOpen
  \bibfield  {author} {\bibinfo {author} {\bibfnamefont {J.}~\bibnamefont {Guo}}, \bibinfo {author} {\bibfnamefont {J.}~\bibnamefont {Tura}}, \bibinfo {author} {\bibfnamefont {Q.}~\bibnamefont {He}},\ and\ \bibinfo {author} {\bibfnamefont {M.}~\bibnamefont {Fadel}},\ }\href {https://doi.org/10.1103/PhysRevLett.131.070201} {\bibfield  {journal} {\bibinfo  {journal} {Physical Review Letters}\ }\textbf {\bibinfo {volume} {131}},\ \bibinfo {pages} {070201} (\bibinfo {year} {2023})}\BibitemShut {NoStop}%
\bibitem [{\citenamefont {Schmied}\ \emph {et~al.}(2016)\citenamefont {Schmied}, \citenamefont {Bancal}, \citenamefont {Allard}, \citenamefont {Fadel}, \citenamefont {Scarani}, \citenamefont {Treutlein},\ and\ \citenamefont {Sangouard}}]{schmiedBellCorrelationsBoseEinstein2016}%
  \BibitemOpen
  \bibfield  {author} {\bibinfo {author} {\bibfnamefont {R.}~\bibnamefont {Schmied}}, \bibinfo {author} {\bibfnamefont {J.-D.}\ \bibnamefont {Bancal}}, \bibinfo {author} {\bibfnamefont {B.}~\bibnamefont {Allard}}, \bibinfo {author} {\bibfnamefont {M.}~\bibnamefont {Fadel}}, \bibinfo {author} {\bibfnamefont {V.}~\bibnamefont {Scarani}}, \bibinfo {author} {\bibfnamefont {P.}~\bibnamefont {Treutlein}},\ and\ \bibinfo {author} {\bibfnamefont {N.}~\bibnamefont {Sangouard}},\ }\href {https://doi.org/10.1126/science.aad8665} {\bibfield  {journal} {\bibinfo  {journal} {Science}\ }\textbf {\bibinfo {volume} {352}},\ \bibinfo {pages} {441} (\bibinfo {year} {2016})}\BibitemShut {NoStop}%
\bibitem [{\citenamefont {Engelsen}\ \emph {et~al.}(2017)\citenamefont {Engelsen}, \citenamefont {Krishnakumar}, \citenamefont {Hosten},\ and\ \citenamefont {Kasevich}}]{engelsenBellCorrelationsSpinSqueezed2017}%
  \BibitemOpen
  \bibfield  {author} {\bibinfo {author} {\bibfnamefont {N.~J.}\ \bibnamefont {Engelsen}}, \bibinfo {author} {\bibfnamefont {R.}~\bibnamefont {Krishnakumar}}, \bibinfo {author} {\bibfnamefont {O.}~\bibnamefont {Hosten}},\ and\ \bibinfo {author} {\bibfnamefont {M.~A.}\ \bibnamefont {Kasevich}},\ }\href {https://doi.org/10.1103/PhysRevLett.118.140401} {\bibfield  {journal} {\bibinfo  {journal} {Physical Review Letters}\ }\textbf {\bibinfo {volume} {118}},\ \bibinfo {pages} {140401} (\bibinfo {year} {2017})}\BibitemShut {NoStop}%
\bibitem [{\citenamefont {Hu}\ and\ \citenamefont {Tura}(2026)}]{huTropicalContractionTensor2026a}%
  \BibitemOpen
  \bibfield  {author} {\bibinfo {author} {\bibfnamefont {M.}~\bibnamefont {Hu}}\ and\ \bibinfo {author} {\bibfnamefont {J.}~\bibnamefont {Tura}},\ }\href {https://doi.org/10.1103/r7s7-c7y2} {\bibfield  {journal} {\bibinfo  {journal} {Physical Review Letters}\ }\textbf {\bibinfo {volume} {136}},\ \bibinfo {pages} {100202} (\bibinfo {year} {2026})}\BibitemShut {NoStop}%
\bibitem [{\citenamefont {Hu}\ \emph {et~al.}(2026)\citenamefont {Hu}, \citenamefont {Vallée}, \citenamefont {Seynnaeve}, \citenamefont {Emonts}, \citenamefont {Mohammadi},\ and\ \citenamefont {Tura}}]{huCharacterizingTranslationinvariantBell2026a}%
  \BibitemOpen
  \bibfield  {author} {\bibinfo {author} {\bibfnamefont {M.}~\bibnamefont {Hu}}, \bibinfo {author} {\bibfnamefont {E.}~\bibnamefont {Vallée}}, \bibinfo {author} {\bibfnamefont {T.}~\bibnamefont {Seynnaeve}}, \bibinfo {author} {\bibfnamefont {P.}~\bibnamefont {Emonts}}, \bibinfo {author} {\bibfnamefont {F.}~\bibnamefont {Mohammadi}},\ and\ \bibinfo {author} {\bibfnamefont {J.}~\bibnamefont {Tura}},\ }\href {https://doi.org/10.1103/mgm4-dz2x} {\bibfield  {journal} {\bibinfo  {journal} {Physical Review A}\ }\textbf {\bibinfo {volume} {113}},\ \bibinfo {pages} {032421} (\bibinfo {year} {2026})}\BibitemShut {NoStop}%
\bibitem [{\citenamefont {Fadel}\ and\ \citenamefont {Tura}(2018)}]{fadelBellCorrelationsFinite2018}%
  \BibitemOpen
  \bibfield  {author} {\bibinfo {author} {\bibfnamefont {M.}~\bibnamefont {Fadel}}\ and\ \bibinfo {author} {\bibfnamefont {J.}~\bibnamefont {Tura}},\ }\href {https://doi.org/10.22331/q-2018-11-19-107} {\bibfield  {journal} {\bibinfo  {journal} {Quantum}\ }\textbf {\bibinfo {volume} {2}},\ \bibinfo {pages} {107} (\bibinfo {year} {2018})}\BibitemShut {NoStop}%
\bibitem [{\citenamefont {Holstein}\ and\ \citenamefont {Primakoff}(1940)}]{holsteinFieldDependenceIntrinsic1940}%
  \BibitemOpen
  \bibfield  {author} {\bibinfo {author} {\bibfnamefont {T.}~\bibnamefont {Holstein}}\ and\ \bibinfo {author} {\bibfnamefont {H.}~\bibnamefont {Primakoff}},\ }\href {https://doi.org/10.1103/PhysRev.58.1098} {\bibfield  {journal} {\bibinfo  {journal} {Physical Review}\ }\textbf {\bibinfo {volume} {58}},\ \bibinfo {pages} {1098} (\bibinfo {year} {1940})}\BibitemShut {NoStop}%
\bibitem [{\citenamefont {Chen}\ \emph {et~al.}(2026)\citenamefont {Chen}, \citenamefont {Hu},\ and\ \citenamefont {Tura}}]{chenOptimizingQuantumViolation2026}%
  \BibitemOpen
  \bibfield  {author} {\bibinfo {author} {\bibfnamefont {J.-F.}\ \bibnamefont {Chen}}, \bibinfo {author} {\bibfnamefont {M.}~\bibnamefont {Hu}},\ and\ \bibinfo {author} {\bibfnamefont {J.}~\bibnamefont {Tura}},\ }\href {https://doi.org/10.48550/arXiv.2511.07523} {\bibinfo {title} {Optimizing quantum violation for multipartite facet {Bell} inequalities}} (\bibinfo {year} {2026}),\ \bibinfo {note} {arXiv:2511.07523}\BibitemShut {NoStop}%
\bibitem [{\citenamefont {Navascués}\ \emph {et~al.}(2007)\citenamefont {Navascués}, \citenamefont {Pironio},\ and\ \citenamefont {Acín}}]{navascuesBoundingSetQuantum2007}%
  \BibitemOpen
  \bibfield  {author} {\bibinfo {author} {\bibfnamefont {M.}~\bibnamefont {Navascués}}, \bibinfo {author} {\bibfnamefont {S.}~\bibnamefont {Pironio}},\ and\ \bibinfo {author} {\bibfnamefont {A.}~\bibnamefont {Acín}},\ }\href {https://doi.org/10.1103/PhysRevLett.98.010401} {\bibfield  {journal} {\bibinfo  {journal} {Physical Review Letters}\ }\textbf {\bibinfo {volume} {98}},\ \bibinfo {pages} {010401} (\bibinfo {year} {2007})}\BibitemShut {NoStop}%
\bibitem [{\citenamefont {Navascués}\ \emph {et~al.}(2008)\citenamefont {Navascués}, \citenamefont {Pironio},\ and\ \citenamefont {Acín}}]{navascuesConvergentHierarchySemidefinite2008}%
  \BibitemOpen
  \bibfield  {author} {\bibinfo {author} {\bibfnamefont {M.}~\bibnamefont {Navascués}}, \bibinfo {author} {\bibfnamefont {S.}~\bibnamefont {Pironio}},\ and\ \bibinfo {author} {\bibfnamefont {A.}~\bibnamefont {Acín}},\ }\href {https://doi.org/10.1088/1367-2630/10/7/073013} {\bibfield  {journal} {\bibinfo  {journal} {New Journal of Physics}\ }\textbf {\bibinfo {volume} {10}},\ \bibinfo {pages} {073013} (\bibinfo {year} {2008})}\BibitemShut {NoStop}%
\bibitem [{\citenamefont {Tavakoli}\ \emph {et~al.}(2024)\citenamefont {Tavakoli}, \citenamefont {Pozas-Kerstjens}, \citenamefont {Brown},\ and\ \citenamefont {Araújo}}]{tavakoliSemidefiniteProgrammingRelaxations2024}%
  \BibitemOpen
  \bibfield  {author} {\bibinfo {author} {\bibfnamefont {A.}~\bibnamefont {Tavakoli}}, \bibinfo {author} {\bibfnamefont {A.}~\bibnamefont {Pozas-Kerstjens}}, \bibinfo {author} {\bibfnamefont {P.}~\bibnamefont {Brown}},\ and\ \bibinfo {author} {\bibfnamefont {M.}~\bibnamefont {Araújo}},\ }\href {https://doi.org/10.1103/RevModPhys.96.045006} {\bibfield  {journal} {\bibinfo  {journal} {Reviews of Modern Physics}\ }\textbf {\bibinfo {volume} {96}},\ \bibinfo {pages} {045006} (\bibinfo {year} {2024})}\BibitemShut {NoStop}%
\bibitem [{\citenamefont {Fadel}\ and\ \citenamefont {Tura}(2017)}]{fadelBoundingSetClassical2017a}%
  \BibitemOpen
  \bibfield  {author} {\bibinfo {author} {\bibfnamefont {M.}~\bibnamefont {Fadel}}\ and\ \bibinfo {author} {\bibfnamefont {J.}~\bibnamefont {Tura}},\ }\href {https://doi.org/10.1103/PhysRevLett.119.230402} {\bibfield  {journal} {\bibinfo  {journal} {Physical Review Letters}\ }\textbf {\bibinfo {volume} {119}},\ \bibinfo {pages} {230402} (\bibinfo {year} {2017})}\BibitemShut {NoStop}%
\bibitem [{\citenamefont {Lipkin}\ \emph {et~al.}(1965)\citenamefont {Lipkin}, \citenamefont {Meshkov},\ and\ \citenamefont {Glick}}]{lipkinValidityManybodyApproximation1965a}%
  \BibitemOpen
  \bibfield  {author} {\bibinfo {author} {\bibfnamefont {H.}~\bibnamefont {Lipkin}}, \bibinfo {author} {\bibfnamefont {N.}~\bibnamefont {Meshkov}},\ and\ \bibinfo {author} {\bibfnamefont {A.}~\bibnamefont {Glick}},\ }\href {https://doi.org/10.1016/0029-5582(65)90862-X} {\bibfield  {journal} {\bibinfo  {journal} {Nuclear Physics}\ }\textbf {\bibinfo {volume} {62}},\ \bibinfo {pages} {188} (\bibinfo {year} {1965})}\BibitemShut {NoStop}%
\bibitem [{\citenamefont {Ma}\ \emph {et~al.}(2011)\citenamefont {Ma}, \citenamefont {Wang}, \citenamefont {Sun},\ and\ \citenamefont {Nori}}]{maQuantumSpinSqueezing2011a}%
  \BibitemOpen
  \bibfield  {author} {\bibinfo {author} {\bibfnamefont {J.}~\bibnamefont {Ma}}, \bibinfo {author} {\bibfnamefont {X.}~\bibnamefont {Wang}}, \bibinfo {author} {\bibfnamefont {C.}~\bibnamefont {Sun}},\ and\ \bibinfo {author} {\bibfnamefont {F.}~\bibnamefont {Nori}},\ }\href {https://doi.org/10.1016/j.physrep.2011.08.003} {\bibfield  {journal} {\bibinfo  {journal} {Physics Reports}\ }\textbf {\bibinfo {volume} {509}},\ \bibinfo {pages} {89} (\bibinfo {year} {2011})}\BibitemShut {NoStop}%
\bibitem [{\citenamefont {Müller-Rigat}\ \emph {et~al.}(2026)\citenamefont {Müller-Rigat}, \citenamefont {Aloy}, \citenamefont {Lewenstein}, \citenamefont {Fadel},\ and\ \citenamefont {Tura}}]{muller-rigatThreeoutcomeMultipartiteBell2026}%
  \BibitemOpen
  \bibfield  {author} {\bibinfo {author} {\bibfnamefont {G.}~\bibnamefont {Müller-Rigat}}, \bibinfo {author} {\bibfnamefont {A.}~\bibnamefont {Aloy}}, \bibinfo {author} {\bibfnamefont {M.}~\bibnamefont {Lewenstein}}, \bibinfo {author} {\bibfnamefont {M.}~\bibnamefont {Fadel}},\ and\ \bibinfo {author} {\bibfnamefont {J.}~\bibnamefont {Tura}},\ }\bibfield  {journal} {\bibinfo  {journal} {npj Quantum Information}\ }\href {https://doi.org/10.1038/s41534-026-01261-8} {10.1038/s41534-026-01261-8} (\bibinfo {year} {2026})\BibitemShut {NoStop}%
\end{thebibliography}%
\end{document}